\documentclass[a4paper,12pt,a4paper]{article}
\pdfoutput=1
\usepackage{packages}
\usepackage{commands}
\begin{document}

\begin{center}
\LARGE{\textbf{Probing higher-spin fields from inflation with higher-order statistics of the CMB}} \\
\end{center}
\vspace{0.3cm}

\begin{center}

\large{Lorenzo Bordin,$^{\rm a}$ Giovanni Cabass $^{\rm b}$}
\\[0.7cm]

\footnotesize{
\textit{$^{\rm a}$}School of Physics \& Astronomy, University of Nottingham, University Park, Nottingham, NG7 2RD, UK}
\\[0.1cm]
\footnotesize{\textit{$^{\rm b}$}Max-Planck-Institut f{\" u}r Astrophysik, Karl-Schwarzschild-Str. 1, 85741 Garching, DE}

\vspace{.1cm}

\end{center}
\vspace{.8cm}

\hrule
\vspace{0.3cm}
\noindent{\small\textbf{Abstract}\\ 
\noindent 
We investigate the degree to which the Cosmic Microwave Background (CMB) can be used to constrain primordial non-Gaussianity 
coming from the presence of spinning particles coupled to the inflaton. We compute the $\langle TTT \rangle$ and $\langle TTTT \rangle$ correlation functions 
arising from the exchange of a particle with spin $s$ and generic mass, and the corresponding signal-to-noise ratios for a cosmic-variance-limited CMB experiment.
We show that already with \emph{Planck} data one could improve the theoretical bounds 
on the amplitude of these primordial templates by an order of magnitude.
We particularly emphasize the fact that the trispectrum could be sizable even if the bispectrum is not, 
making it a prime observable to explore the particle content during inflation.} 
\vspace{0.3cm}
\noindent
\hrule
\vspace{.8cm}

\tableofcontents

\section{Introduction}

\normalsize
Originally introduced to solve the problems of the Hot Big Bang cosmology, the inflationary paradigm owes its success to how naturally it provides the initial conditions of our Universe. 
In the simplest scenario, a scalar field with a time-dependent vacuum expectation value (v.e.v.), the inflaton, 
is responsible for the accelerated expansion of the universe and its quantum fluctuations, stretched up to cosmological scales, 
seed the Cosmic Microwave Background (CMB) anisotropies and the distribution of galaxies in the Large-Scale Structure (LSS). 
In this scenario only two additional degrees of freedom are active, besides the scalar excitations of the inflaton: these are the tensor helicities of the graviton. 

Despite the minimal model of inflation is enough to describe the current observations, additional degrees of freedom could have been active during the inflationary epoch. 
This is especially true if inflation happened at energies close to the current \emph{Planck} upper bound $H\lesssim\SI{6e13}{\mathrm{GeV}}$ \cite{Akrami:2018odb}. 
In fact, every particle with mass $m\lesssim H$ gets excited during inflation and might then contribute to the primordial correlation functions.

Understanding the particle content during the inflationary epoch is one of the hardest and most exciting challenges of modern cosmology. 
Much work has been done to characterize the non-Gaussian signatures due to presence of additional light scalar fields. 
However, until very recently, less effort has been devoted to understanding the effects of particles with a higher spin. 
One of the main reasons for this consists in the difficulty of having light perturbations with spin $s>1$ in a de Sitter spacetime (that well approximates the inflationary spacetime). 
In fact, the behaviour of spinning fields in de Sitter is very constrained by the de Sitter isometries, that force particles to obey the so called Higuchi Bound \cite{HIGUCHI1987397} 
\be
m^2 > s(s-1) H^2\,\,.
\ee
Close to the boundary of de Sitter, \ie~when the conformal time $\eta$ goes to $0$, its isometries also fix the time evolution of fields in terms of their masses and spins,
\be
\label{eq:AHM_definition}
\s_{i_1, \dots , i_s} (\eta,\k) \sim ({-\eta})^{\Delta_s} \O_{i_1, \dots , i_s}(\k)\,\,, \qquad \Delta_s = 3/2 - \sqrt{(s-1/2)^2-(m/H)^2}
\ee 
(see for example \cite{Arkani-Hamed:2015bza}). 
Combined with the Higuchi Bound, the above time evolution implies that particles with spin decay exponentially fast during inflation, 
suppressing the contribution they might add to the primordial correlators.

The difficulty of having light fields during inflation can be overcome by coupling higher-spins states with the foliation provided by the inflaton.\footnote{The coupling 
with the foliation is not the only way to have light spinning particles. 
Another possibility consists in having fields that enjoy a partial gauge invariance that allows to evade the Higuchi Bound. 
These fields, called partially massless states, have unsuppressed super-horizon perturbations for some values of their mass \cite{Baumann:2017jvh,Franciolini:2017ktv}.} 
The time dependent v.e.v. of the inflaton breaks the \emph{special conformal transformations}, the three de Sitter isometries responsible for the Higuchi Bound, 
and it can then make light spinning particles healthy \cite{Bordin:2016ruc} (see also \cite{Kehagias:2017cym}). 
These higher-spin perturbations, coupled with the inflaton, can be described using an effective approach which extends the formalism of the 
\emph{effective field theory of inflation} (EFTI) \cite{Cheung:2007st,Cheung:2007sv}. 
This formalism allows to study additional perturbations with arbitrary spin \cite{Bordin:2018pca}.\footnote{In the language of the 
Effective Field Theory these higher-spin states should be viewed as excitations of a fluid more than elementary particles.} 

The presence of long-lived perturbations with spin $s>1$ generates some anisotropy that does not decay even on the longest scales.
It manifests itself in the non-Gaussian $3$- and $4$-point functions of scalar perturbations, giving rise to new shapes of scalar non-Gaussianity. 
These new shapes are quite different from the templates of non-Gaussianity available on the market, namely the local, the equilateral, and the orthogonal templates. 
For instance, while the scalar bispectrum and trispectrum peak in the squeezed ($k_{1} \ll k_2 \simeq k_3$) 
and counter-collapsed ($k_{12} \ll k_1, k_3$, $k_1 \simeq k_2, k_3\simeq k_4$) configurations respectively, 
like the common templates for local and $\tau_{\rm NL}$-type non-Gaussianity, their scaling with the momenta is different from that of these parameterizations. 
Indeed, it depends on the mass of the exchanged light field, and could even be non-analytic.
Moreover, in these configurations both correlation functions are modulated by the angle between long and short modes, 
with the modulation being a function of the spin of the exchanged particle. 
For these reasons, these new shapes of non-Gaussianity show a very small overlap with the {local} template and the $\tau_{\rm NL}$ parametrization of the trispectrum, 
which are the only templates constrained by \emph{Planck} data that peak in the squeezed and counter-collapsed configurations respectively. 
Moreover the non-Gaussianity generated by the exchange of higher-spin fields cannot be mimicked by single-field inflation or 
by multi-field models with only scalar fields: it therefore represents a smoking gun for the presence of higher-spin excitations. 
All this suggests that a dedicated CMB analysis of the \emph{Planck} data, aimed at constraining these shapes of non-Gaussianity, is required. 

In this paper we take a first step in this direction with a Fisher analysis of how a cosmic-variance-limited experiment 
measuring temperature anisotropies of the cosmic microwave background up to $\ell_{\rm max} = 3500$ can constrain the templates
\begin{align}
&\avg{\zeta_{\k_1} \zeta_{\k_2} \zeta_{\k_3}}' = 
\F_s \(\frac{k_1 k_2 k_3}{k_{\rm t}^3/8}\)^{\Delta} P_\zeta(k_1) P_\zeta(k_2) \LP_s(\vers{k}_1 \cdot \vers{k}_2) + \text{$2$ perms.}\,\,, \label{bis_temp} \\
&\avg{\zeta_{\k_1} \zeta_{\k_2} \zeta_{\k_3} \zeta_{\k_4}}' =
\tau_s \[ \(\frac{k_{12}^2}{k_1 k_3}\)^\Delta P_{\zeta}(k_{12}) P_\zeta(k_1) P_{\zeta}(k_3) {\rm P}_s(\vers{k}_1 \cdot\vers{k}_3) + \text{$23$ perms.}\]\,\,, \label{tris_temp}
\end{align}
where the prime means that we have removed the factor $(2\pi)^3\delta^{(3)}(\vec{k}_{12\dots})$, and we define $k_{\rm t} = k_1 + k_2 + k_3$ and $k_{ij} = \abs{\k_i+\k_j}$.
These templates parametrize well the effects of a particle $\s$ with spin $s$ and mass $m$. 
The function $\LP_s (x)$ is the Legendre polynomial of order $s$, while from now on we define\footnote{Notice that, 
differently from Eq.~\eqref{eq:AHM_definition}, this $\Delta$ does \emph{not} depend on $s$. 
This redefinition does not lead to any loss of generality: we refer to, e.g., Ref.~\cite{Bordin:2018pca} for details.} 
\begin{equation}
\label{eq:capital_delta_definition}
\Delta = 3/2 - \sqrt{9/4 - (m/H)^2}\,\,.
\end{equation}
Notice that in the case of an additional massless scalar, \ie~if $s=0$ and $\Delta=3/2$, the two templates reduce to the well-known parameterizations 
$\F_0 = f_{\rm NL}$ and $\tau_0 = \tau_{\rm NL}$. 

There are several works in the literature that study the observational consequences higher-spins fields in the CMB. 
For instance, Refs.~\cite{Shiraishi:2013vja,Franciolini:2018eno} analysed the bispectrum template, 
Eq.~\eqref{bis_temp}, for $\Delta=0$, i.e.~when the exchanged spin-$s$ particle is massless. 
Moreover, Ref.~\cite{Shiraishi:2013oqa} has studied a template similar to that of Eq.~\eqref{tris_temp} 
for $s=2$ and $\Delta=0$,\footnote{This template is different from Eq.~\eqref{tris_temp}, 
both in its angular modulation and scale dependence (for $\Delta\neq 0$) in the counter-collapsed limit.} 
while Ref.~\cite{Bartolo:2017sbu} studied the effects of primordial higher-spin fields on the power spectrum 
(see also Ref.~\cite{Bartolo:2012sd} for an early discussion). 
In this paper we make a further step, implementing also the scale dependence that arises when the additional degrees of freedom are not massless 
and extending the trispectrum analysis even to particles with spin $s>2$. 

Many authors have also studied the effects of primordial higher-spin fields on LSS observables, 
mainly focusing on the primordial bispectrum. An incomplete list of works on this topic is 
\cite{Assassi:2015jqa,Schmidt:2015xka,Raccanelli:2015oma,Assassi:2015fma,Chisari:2016xki,MoradinezhadDizgah:2018ssw,MoradinezhadDizgah:2018pfo}. 

Our paper is organized as follows. 
In Section~\ref{theory_sec} we briefly review how the templates of Eqs.~\eqref{bis_temp},~\eqref{tris_temp} arise in theories where additional light spinning particles are active. 
In Section~\ref{cmb_signal_sec} we derive the expected signal in the $3$- and $4$-point functions of CMB temperatures anisotropies. 
Section~\ref{analysis_sec} is devoted to the forecast for cosmic-variance-limited CMB experiments. We comment the results and conclude in Section~\ref{discussion_sec}. 
Appendices \ref{appendix-A} to \ref{appendix-D} collect some technical details. 

\paragraph{Notation and conventions.}
We denote the modulus of a vector by $\abs{\k}\equiv k$. We then use the following shorthand notation: $\k_{ij} \equiv \k_i + \k_j$, $L_{ij} \equiv L_i+L_j$, and so on. 
We denote $\sum_{i=1}^3\vec{k}_i$ by $\k_{123}$ and $\sum_{i=1}^4\vec{k}_i$ by $\k_{1234}$. 
The functions $\Y \ell m$, $\Ys \ell m$ are, respectively, the spherical harmonic of index $\ell,m$ and its complex conjugate. 
The functions ($\LP_\ell^m$) $\LP_\ell$ are the (associated) Legendre polynomials, while $j_\ell$ are the spherical Bessel functions of the first kind. 
We consider a flat $\Lambda$CDM cosmology compatible with the latest \emph{Planck} results \cite{Ade:2015xua}: 
$A_{\rm s} = \num{2.1e-9}$, $n_{\rm s}=\num{0.965}$, $k_\ast = \SI{0.05}{\mathrm{Mpc}^{-1}}$, 
$r=0$, $\Omega_bh^2=\num{0.0226}$, $\Omega_ch^2=0.112$, $\sum m_\nu=\SI{0.06}{\mathrm{eV}}$, $H_0=\SI{67.5}{\mathrm{km}.\mathrm{s}^{-1}\mathrm{Mpc}^{-1}}$, 
$T_{\rm CMB}=\SI{2.7255}{\mathrm{K}}$, $Y_{\rm He}=0.24$, $N_{\rm eff}=3.046$, $\tau=0.06$. 
The various transfer functions for temperature anisotropies are computed with \texttt{CAMB} \cite{Lewis:1999bs}, version 0.1.8.1.\footnote{\url{https://camb.info/readme.html}.} 
We use the \texttt{wigxjpf} library (version 1.9) to compute the necessary Wigner symbols \cite{Johansson:2015cca}.\footnote{\url{http://fy.chalmers.se/subatom/wigxjpf/}.}

\section{The primordial signal}
\label{theory_sec}

The existence of light excitations with spin $s\geq2$ during inflation, in a quasi-de Sitter spacetime, 
is made possible by the coupling with the preferred foliation that breaks some of the de Sitter isometries. 
Since we are interested in working out the phenomenological consequences of these higher-spin states, 
it is convenient to use an effective approach to describe the additional degrees of freedom. 
The formalism presented in \cite{Bordin:2018pca} is well-suited to describe generic spin-$s$ excitations which live on the hypersurfaces of constant inflaton. 

Because of the foliation, Lorentz invariance is broken and thus on sub-Hubble scales fields are invariant only under spatial rotations. 
This implies that one should cast perturbations into representations of $SO(3)$ instead of $SO(1,3)$. 
Spin-$s$ excitations are therefore described by traceless rank-$s$ tensors $\Sigma^{i_1\dots i_s}$ 
that live on the three-dimensional surfaces at constant inflaton, $\psi \equiv t + \pi(t,\vec{x}) = \text{const.}$ 
Diffeomorphism invariance is then restored in the action by ``pushing forward'' the tensor $\Sigma^{i_1\dots i_s}$ 
in a way that depends on the particular configuration of the inflaton slices described by $\pi$ 
\be
\Sigma^{\mu_1 \dots \mu_s} (\Sigma^{i_1\dots i_s}, \pi) = \Sigma^{i_1\dots i_s} 
\left. \frac{\d x^{\mu_1}}{\d x^{i_1}}\right|_\psi \ \dots \ \left. \frac{\d x^{\mu_s}}{\d x^{i_s}}\right |_\psi\,\,.
\ee
The generic action for a spin-$s$ field that preserves spatial diffeomorphisms 
can be written using $\Sigma^{\mu_1 \dots \mu_s}$ and the unit vector $n^\mu$ perpendicular to the inflaton hypersurfaces.
At quadratic level in $\Sigma$ one gets
\begin{equation}
\label{eq:spin_s_quad_action}
\begin{split}
&S = \frac1{2 s!} \int{\rm d} t{\rm d}^3x\,a^3\,\Big( (1- c_s^2) \, n^\mu n^\lambda \, \nabla_\mu 
\Sigma^{\nu_1 \dots \nu_s} \nabla_\lambda \Sigma_{\nu_1 \dots \nu_s} - \, c_s^2 \,\nabla_\mu \Sigma^{\nu_1 \dots \nu_s} \nabla^\mu \Sigma_{\nu_1 \dots \nu_s} \\
&\hphantom{S = \frac1{2 s!} \int{\rm d} t{\rm d}^3x\,a^3\,\Big( } - \delta c_s^2 \, \nabla_\mu 
\Sigma^{\mu\nu_2 \dots \nu_s} \nabla_\lambda {\Sigma^\lambda}_{\nu_2 \dots \nu_s} - (m^2 + s \, c_s^2 H^2) \, \Sigma^{\nu_1 \dots \nu_s} \Sigma_{\nu_1 \dots \nu_s}\Big) \\
&\hphantom{S} = 
\frac1{2 s!} \int {\rm d} t {\rm d}^3x\,a^3 \,\Big( (\dot \sigma^{i_1 \dots i_s})^2 - \, c_s^2 \, a^{-2} (\d_j \sigma^{i_1 \dots i_s})^2 - \, \delta c_s^2 \, a^{-2} (\d_j \sigma^{j i_2 \dots i_s})^2
- \, m^2 \, (\sigma^{i_1 \dots i_s})^2 \Big)\,\,,
\end{split}
\end{equation}
where, in the last line, we have introduced a new field $\sigma^{i_1 \dots i_s} \equiv a^s \, \Sigma^{i_1 \dots i_s}$, 
which has the same temporal part of the kinetic term as a canonical scalar field with mass $m^2$. 
Notice that there are three independent kinetic terms one can write. 
This means that different helicities have a different propagation speed $c^2_h$ (for $h=0,\dots,s$): 
this is a function of the speed of propagation of the helicity-$s$ mode, $c_s^2$, and of the parameter $\delta c_s^2$. 

For a systematic study of the phenomenology of the spin-$s$ fields one should also include interaction and mixing terms allowed by the symmetries. 
Even if we work at leading order in fields and derivatives, it is hard to write the most generic operators for generic spin $s$. 
For simplicity, we just report the leading operators in the case of an additional spin-$2$ particle, \ie~
\be
\label{int_action}
S_{{ \rm int}} = \int {\rm d}^4x\,\sqrt{-g}\,\Big( \mpl \, \rho \, \delta K_{\alpha\beta} \Sigma^{\alpha\beta} + \mpl \,\widetilde{\rho} \, \delta g^{00} 
\delta K_{\alpha\beta} \Sigma^{\alpha\beta} - \mu \, \Sigma^{\alpha\beta} {\Sigma_\alpha}^\gamma \Sigma_{\gamma\beta} \Big) \,\,,
\ee
where $\rho$, $\widetilde{\rho}$ and $\mu$ are coupling constants and $\delta K_{\alpha\beta} \equiv K_{\alpha\beta} - a^2 H h_{\alpha\beta}$ 
is the fluctuation of the extrinsic curvature of constant-$\psi$ hypersurfaces. Notice that the first term of the above action starts quadratic in perturbations: 
it mixes $\s$ with the inflaton field. The second and the third terms, instead, start cubic: at this order in perturbations they give rise to, respectively, a $\s \pi \pi$ and a $\s^3$ vertex.

\paragraph{Non-Gaussian signal.}
The mixing and interaction terms of $\s$ with the inflaton perturbations $\pi$ generate complicate momentum dependencies in the scalar bispectrum and trispectrum. 
However, these dramatically simplify in the squeezed and counter-collapsed limit, respectively. 
Taking the limit $k_1 \ll k_2 \simeq k_3$ the bispectrum becomes 
\be
\label{bispectrum_spin-s_exchng}
\avg{\zeta_{\k_1} \zeta_{\k_2} \zeta_{\k_3}} = (2\pi)^3 \delta^{(3)}(\k_{123}) \ 
\frac{\widetilde{\F}_s}{c_0^{2\nu}} \(\frac{k_1}{k_2}\)^{\Delta} P_\zeta(k_1) P_\zeta(k_2) \(\ep^{(0)}_{s, \ i_1 \dots i_s}(\hat{\vec{k}}_{1})\,\hat{k}_{2\,i_1}\dots \hat{k}_{2i_s}\)\,\,,
\ee
while in the limit $k_{12} \ll k_1, k_3$ and $k_1 \simeq k_2, k_3\simeq k_4$, the trispectrum simplifies into 
\be
\label{trispectrum_spin-s_exchng}
\begin{split}
\avg{\zeta_{\k_1} \zeta_{\k_2} \zeta_{\k_3} \zeta_{\k_4}} = (2\pi)^3 \delta^{(3)}(\k_{1234}) &
\ \widetilde{\tau}_s \(\frac{k_{12}^2}{k_1 \, k_3}\)^{\Delta} P_{\zeta}(k_{12}) P_\zeta(k_1) P_{\zeta}(k_3)\,\times \\
& \sum_{h=-s}^{s} c_h^{-2\nu} \(\ep^{(h)}_{s, \ i_1 \dots i_s}(\hat{\k}_{12})\,\hat k_{1\,i_1} \dots \hat k_{1\,i_s} \)\,\times \\
&\hphantom{ \sum_{h=-s}^{s} c_h^{-2\nu}}\, \(\ep^{(h)}_{s, \ j_1 \dots j_s}(\hat{\k}_{12})\,\hat k_{3\,j_1} \dots \hat k_{3\,j_s} \)^*\,\,.
\end{split}
\ee
In the above formulae, $c_h$ is the speed of propagation of the helicity-$h$ component of the particle $\s$, 
$\Delta$ is $3/2-\nu$ (with $\nu$ related to the mass of the particle, $\nu = \sqrt{9/4-(m/H)^2}$), 
and finally $\ep^{(h)}_{s, \ j_1 \dots j_s}(\vers k)$ is the polarization tensor of the helicity-$h$ component of $\s$.

\begin{figure}[h]
\centering
\begin{tabular}{c c}
\includegraphics[width=0.45\columnwidth]{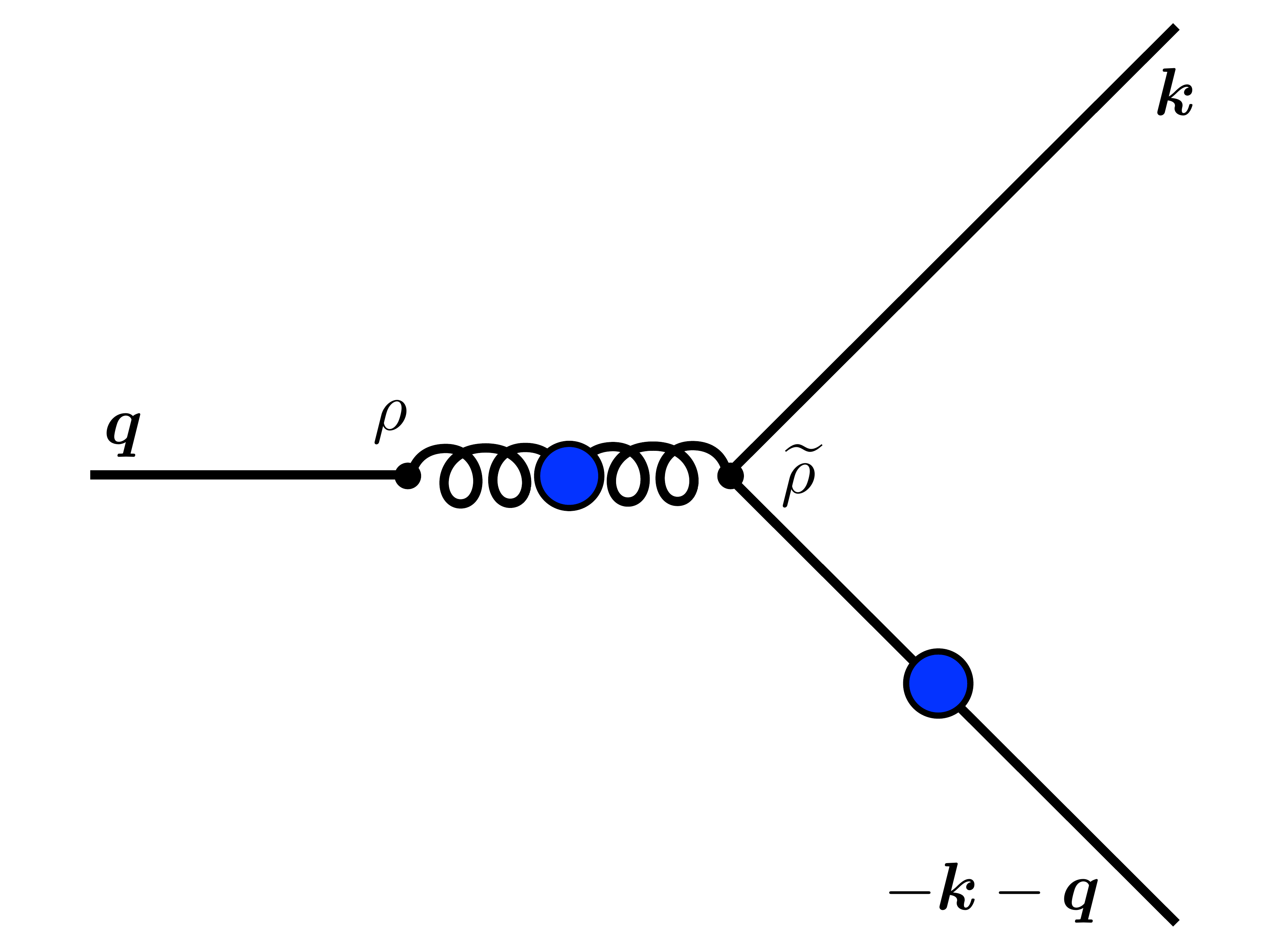} \quad\quad 
\includegraphics[width=0.45\columnwidth]{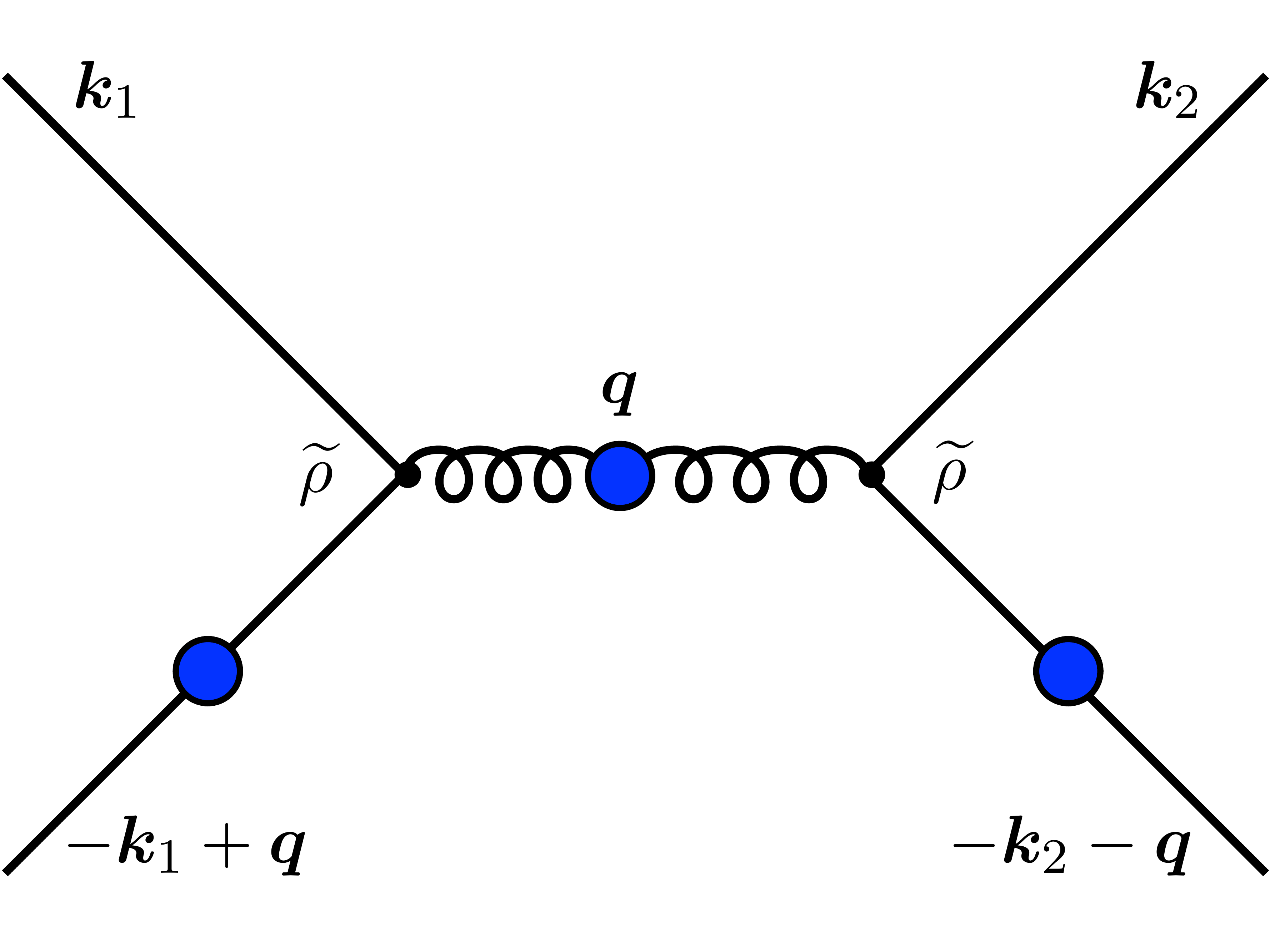}
\end{tabular}
\caption{Leading contributions to the scalar $3$- and $4$-point functions. 
Solid lines correspond to $\pi$, curly lines to $\sigma$. The blue circles indicate a contraction between a pair of free fields, \ie~the insertion of a power spectrum. 
One has to put the minimal number of circles in such a way that external lines cannot be connected without going through a circle (contraction), 
and each circle is connected to external lines from both sides. Diagrams where the circles are located in different places are subdominant for \mbox{small $c_h$.}} 
\label{fig_corr_functions}
\end{figure}

Having in mind the example of the additional spin-$2$ particle, Eq.~\eqref{int_action}, 
we can estimate how the coefficients $\F_s$ and $\tau_s$ are related to the coupling constants of the quadratic and cubic interactions. 
In terms of the canonically normalized scalar perturbations $\pi_{\rm c}$ the mixing is schematically of the form $\sim \rho /(\sqrt \ep H)\, \d \d \pi_{\rm c} \,\sigma $, 
while the $\pi\pi\s$ vertex is $\sim \widetilde{\rho} / (\epsilon H^2 \mpl)\, \dot \pi_{\rm c}\, \d \d\pi_{\rm c}\, \s$,\footnote{We have neglected, 
for the sake of simplicity, the cubic vertex proportional to $\rho$. } therefore one gets (see Fig.~\ref{fig_corr_functions})
\begin{align}
& \F_s \equiv \frac{\widetilde{\F}_s}{c_0^{2\nu}} \sim \( \frac{\rho}{H \sqrt \epsilon} \) \( \frac{\widetilde{\rho} }{H \sqrt \epsilon}\) \frac{1}{c_0^{2\nu}} \,\,, \label{estimates_3} \\
& \tau_s \equiv \frac{\widetilde{\tau}_s}{c_h^{2\nu}}\sim \(\frac{\widetilde{\rho}}{H \sqrt \epsilon} \)\(\frac{\widetilde{\rho}}{H \sqrt \epsilon} \) \frac{1}{c_h^{2\nu}}\,\,, \label{estimates_4}
\end{align}
where the $c_h$ stands for the speed of propagation of the helicity exchanged in the horizontal propagator.
Let us make a few comments on the amplitude of $\F_s$ and $\tau_s$.
First, looking at the diagram on the left of Fig.~\ref{fig_corr_functions}, we immediately realize that the scalar bispectrum 
can get enhanced only if the helicity-$0$ component of $\s$ mixes with the inflaton. 
In fact, at linear order in perturbations, scalars cannot mix with fields of a different helicity.
On the other hand, the scalar trispectrum is non-vanishing even if there is no linear mixing with the inflaton.
Furthermore, even in presence of such linear mixing, if $c_0\simeq 1$ but $c_{1,2} \ll 1$, the $4$-point function will still be enhanced with respect to the $3$-point function. 
For these reasons it is important to analyze both correlation functions. 
It is possible to get rough bounds on the maximum amplitude of $\F_s$ and $\tau_s$ by studying the theory described by Eqs.~\eqref{eq:spin_s_quad_action},~\eqref{int_action}. 
The requirements $\rho, \widetilde{\rho} \ll \sqrt \epsilon H$ ensure that the theory is in the perturbative regime and that there are no gradient instabilities. 
At the same time, to satisfy observational constraints, we require $c_s \geq 10^{-2}$ (we refer to \cite{Bordin:2018pca} for details). 
From the estimates of Eqs.~\eqref{estimates_3},~\eqref{estimates_4} we therefore see that the amplitude of these new shape of non-Gaussianity could be quite sizable.

\paragraph{Primordial Templates.}
Eq.~\eqref{bispectrum_spin-s_exchng} is well approximated by the template of Eq.~\eqref{bis_temp}, \ie~
\be
\label{bispectrum_templ_full}
\avg{\zeta_{\k_1} \zeta_{\k_2} \zeta_{\k_3}} = (2\pi)^3 \delta(\k_{123})\ \F_s \(\frac{k_1 k_2 k_3}{k_{\rm t}^3/8}\)^{\Delta} P_\zeta(k_1) 
P_\zeta(k_2) \LP_s(\hat{\vec{k}}_1 \cdot \hat{\vec{k}}_2) + \text{$2$ perms.} \,\,,
\ee
where $k_{\rm t} = k_1 + k_2 + k_3$. 
However, this shape is too difficult to handle because it is not separable. 
Rather than pursuing the analysis of this shape we will use its squeezed-limit approximation. 
As it is shown in Appendix~\ref{appendix-A}, we are allowed to do this since most of the signal comes from the squeezed configuration for $0 < \Delta\leq 1$. 
We therefore use the following template for the primordial bispectrum 
\begin{equation}
\label{eq:bispectrum_templ}
\begin{split}
\avg{\zeta_{\k_1} \zeta_{\k_2} \zeta_{\k_3}} &= 
(2\pi)^3 \delta(\k_{123}) \ \F_s \(\frac{k_1}{k_2}\)^{\Delta} P_\zeta(k_1) P_\zeta(k_2) \LP_s(\hat{\vec{k}}_1 \cdot \vers{k}_2) + (\vec{k}_2 \to \vec{k}_3) \\
&\equiv (2\pi)^3\delta(\k_{123}) \ \F_s\, b_{\k_1\k_2\k_3}(s) + (\vec{k}_2 \to \vec{k}_3)\,\,,
\end{split}
\end{equation}
with
\be
\label{eq:def_b_k1_k2_k3}
b_{\k_1\k_2\k_3}(s) = \frac{4\pi}{2s+1} \(\frac{k_1}{k_2}\)^{\Delta} P_\zeta(k_1) P_\zeta(k_2) \, \sum_{h=-s}^{s} \Y sh (\vers k_1) \Ys sh (\vers k_2) 
+ (\vec{k}_2 \to \vec{k}_3)\,\,,
\ee
where we have expressed the contractions between the wavevectors and the polarization tensor first in terms of the Legendre Polynomials $\LP_s(x)$ 
and then in terms of the Spherical Harmonics $\Y \ell m (\theta,\phi)$. 

Let us move now to the trispectrum. Again, we first need to rewrite the polarization tensors in terms of the spherical harmonics (see \eg~Ref.~\cite{Baumann:2017jvh}). 
First of all we rewrite the second line of Eq.~\eqref{trispectrum_spin-s_exchng} as 
\begin{equation}
\label{eq:general_modulation}
\begin{split}
\(\ep^{(h)}_{s, \ i_1 \dots i_s}(\hat{\k}_{12})\,\hat k_{1\,i_1} \dots \hat k_{1\,i_s} \) 
\(\ep^{(h)}_{s, \ j_1 \dots j_s}(\hat{\k}_{12})\,\hat k_{3\,j_1} \dots \hat k_{3\,j_s} \)^* &= \Y s h(\theta,\phi)\, \Ys s h(\theta',\phi') \\
&= e^{{\rm i} h \psi} \, { \LP}_s^h(\cos \theta) \, {\LP}_s^h(\cos \theta')\,\,,
\end{split}
\end{equation}
where $\theta$ ($\theta'$) is the angle between $\vers k_{12}$ and $\vers k_1$ ($\vers k_3$), 
$\phi$ ($\phi'$) is the angle of the projection of $\vers k_1$ ($\vers k_3$) on the plane perpendicular to $\vers k_{12}$ 
and, finally, $\psi = \phi-\phi'$. A further simplification is needed to be able to compute the CMB $4$-point function. 
This is because the right-hand side of the above expression makes a precise choice of $\vers k_{12}$, 
while we need to integrate over all possible directions $\vers k_{12}$ in order to compute the signal in multipole space (see Section~\ref{cmb_signal_sec}). 
Therefore, we further assume that all the helicities of $\s$ propagate with same speed, \ie~$c_h = c_s\ \forall\ h = 0, \dots, s$. 
With this assumption they all contribute with the same amplitude to the primordial trispectrum, 
and the second line of Eq.~\eqref{trispectrum_spin-s_exchng} becomes 
\be
c_s^{-2\nu} \sum_h \Y s h(\theta, \phi) \Ys s h (\theta',\phi') \propto c_s^{-2\nu} 
\, \LP_s( \vers k_1 \cdot \vers k_3) = \frac{4\pi}{2s+1} \ \sum_{h=-s}^s\, \Y sh (\vers k_1) \, \Ys sh (\vers k_3)\,\,.
\ee
With this assumption the angular modulation does no longer depend on the angle $\psi$ (in other words, we got rid of the $\vers k_{12}$ dependence). 
In conclusion, we use as template the following expression
\begin{equation}
\label{eq:trispectrum_templ}
\begin{split}
\avg{\zeta_{\k_1} \zeta_{\k_2} \zeta_{\k_3} \zeta_{\k_4}} &= (2\pi)^3 \delta^{(3)}(\k_{1234})\ \tau_s \[ \LP_s(\vers k_1 \cdot \vers k_3) 
\(\frac{k_{12}^2}{k_1 k_3}\)^\Delta P_{\zeta}(k_{12}) P_\zeta(k_1) P_{\zeta}(k_3)+\text{$23$ perms.}\] \\
&\equiv (2\pi)^3 \, \tau_s \int{\rm d}^3K \, \delta^{(3)}(\k_{12}-\vec K) \,\delta^{(3)}(\k_{34} + \vec K) \ t\,^{\k_1 \k_2}_{\k_3 \k_4}(K,s) + \text{$23$ perms.}\,\,,
\end{split}
\end{equation}
with 
\be
t\,^{\k_1 \k_2}_{\k_3 \k_4}(K,s) = \frac{4\pi}{2s+1} \(\frac{K^2}{k_1 k_3}\)^\Delta P_\zeta(K) P_\zeta(k_1) P_\zeta(k_3) \sum_{h=-s}^s \Y sh(\vers k_1) \,\Ys sh (\vers k_3)\,\,.
\ee

\section{Signal in the sky}
\label{cmb_signal_sec}

With Eqs.~\eqref{eq:bispectrum_templ},~\eqref{eq:trispectrum_templ} at hand we can move to the computation of the CMB correlation functions. 
It is convenient to start by writing the primordial templates in multipole space, using the variable $\zeta_{\ell m} = \int {\rm d}^2\hat k\,\Ys \ell m (\vers k) \zeta_{\k}$. 
We begin with the computation of the bispectrum, \ie~from Eq.~\eqref{eq:bispectrum_templ}. In terms of $\zeta_{\ell m}(k)$ it reads
\be
\begin{split}
\avg{\prod_{i=1}^3\zeta_{\ell_im_i}(k_i)} &= \int \prod_{i=1}^3 {\rm d}^2 \hat k_i \, \Ys {\ell_i}{m_i}(\vers k_i) \, \avg{\zeta_{\k_1} \zeta_{\k_2} \zeta_{\k_3}} \\
&=\F_s \int{\rm d}^3x\, \[\prod_i {\rm d}^2 \hat k_i \ \Ys{\ell_i}{m_i}(\vers k_i)\] e^{{\rm i} \k_{123} \cdot \x}\,b_{\k_1\k_2\k_3}(s) + (\vec{k}_2 \to \vec{k}_3)\,\,, 
\end{split}
\ee
where in second line we have used the integral expression of the Dirac delta. We then use the Rayleigh expansion of the exponentials, \ie~
\be
e^{{\rm i}\k\cdot\x} = 4\pi \sum_{\ell,m} \, {\rm i}^\ell \, j_\ell(kx) \, \Y \ell m (\vers k) \, \Ys \ell m (\vers x)\,\,,
\ee
and we perform the integrals over all the possible directions $\vers k_i$ exploiting the orthonormality of the spherical harmonics and the Gaunt integral, 
\be
\G^{\ell_1 \,\ell_2\, \ell_3}_{m_1 m_2 m_3} = \int {\rm d}\Omega \, \Y{\ell_1}{m_1}(\vers n)\Y{\ell_2}{m_2}(\vers n)\Y{\ell_3}{m_3}(\vers n) = h^{\ell_1\,\ell_2\,\ell_3} \times
{
\begin{pmatrix} 
\ell_1 \! & \! \ell_2 \! & \! \ell_3 \\
m_1 \! & \! m_2 \! & \! m_3 
\end{pmatrix}
}
\,\,,
\ee
where
\be
h^{\ell_1\,\ell_2\,\ell_3} = \sqrt{\frac{(2\ell_1+1)(2\ell_2+1)(2\ell_3+1)}{4\pi}} \, 
{
\begin{pmatrix} 
\ell_1 \! & \!\ell_2 & \! \ell_3 \\
 0 \! & \! 0 \! & \! 0 
\end{pmatrix}
}
\,\,.
\ee
We finally obtain
\be
\begin{split}
\avg{\prod_{i=1}^3\zeta_{\ell_im_i}(k_i)} &= \frac{4\pi \F_s}{2s+1} \int_0^{\infty}{\rm d}x \, x^2 \, k_1^\Delta j_{L_1}(k_1 x) P_\zeta(k_1) \, k_2^{-\Delta} 
j_{L_2}(k_2 x) P_\zeta(k_2) \, j_{L_3}(k_3 x) \,\times \\
&\hphantom{= \frac{4\pi \F_s}{2s+1} \int_0^{\infty}{\rm d}x \, x^2\,} 
\sum_{L_1,L_2, M_1,M_2} \sum_h {\rm i}^{L_{12}+\ell_3} (-1)^{h+m_3} \ \G^{\ell_1\, L_1\, s}_{m_1M_1h} \ \G^{\ell_2\, L_2\, s}_{m_2M_2h} \ \G^{L_1\, L_2\, \ell_3}_{M_1M_2m_3} \\
&\;\;\;\; + (\vec{k}_2 \to \vec{k}_3)\,\,.
\end{split}
\ee
Using the definition of Eq.~\eqref{eq:def_b_k1_k2_k3}, the above expression can be written in a way that is manifestly isotropic in terms of the Wigner $6$-${\rm j}$ symbols, \ie~
\be
\label{primordial_bispectrum_multipole_sp}
\avg{\prod_{i=1}^3\zeta_{\ell_im_i}(k_i)} = 
{
\begin{pmatrix} 
\ell_1 \!\! & \!\! \ell_2 \!\! & \!\! \ell_3 \\
-m_1 \!\! & \!\! -m_2 \!\! & \!\! m_3 
\end{pmatrix}
}
\, (2\pi)^3 \F_s \ b_{k_1 \, k_2 \, k_3}^{\ell_1 \, \ell_2 \, \ell_3} (s) + (\vec{k}_2 \to \vec{k}_3)\,\,,
\ee
where
\be
\begin{split}
\label{reduced_bispectrum_multipole_space}
b_{k_1 \, k_2 \, k_3}^{\ell_1 \, \ell_2 \, \ell_3} (s)= \frac{32 \pi}{2s+1} \sum_{L_1,L_2} &\,I^{\ell_1\, \ell_2\, \ell_3}(L_1,L_2,s)\,\times \\ 
& \int_0^{\infty}{\rm d}x \, x^2 \, k_1^\Delta j_{L_1}(k_1 x) P_\zeta(k_1) \, k_2^{-\Delta} j_{L_2}(k_2 x) P_\zeta(k_2) \, j_{L_3}(k_3 x)\,\,.
\end{split}
\ee
The symbol $I$ encodes the angular structure of the primordial template and it is given by 
\be
I^{\ell_1\,\ell_2\,\ell_3}(L_1,L_2,s) = (-1)^{\frac{\ell_1}{2}+\frac{3}{2}\ell_2+\ell_3+\frac{3}{2}L_1+\frac{L_2}{2}} \ h^{L_1\,L_2\,\ell_3} \ h^{\ell_1\,L_1\,s} \ h^{\ell_2\,L_2\,s} \
{
\begin{Bmatrix} 
\ell_1 \! & \! \ell_2 \! & \! \ell_3 \\
L_2 \! & \! L_1 \! & \! s 
\end{Bmatrix}\,\,,
}
\ee
where the curly matrices stand for the Wigner $6$-${\rm j}$ symbols.

Let us move now to the trispectrum template, \ie~Eq.~\eqref{eq:trispectrum_templ}. 
The computation is similar to that of the bispectrum template. In terms of $\zeta_{\ell m}(k)$ we have
\be
\label{trispectrum_multipole_space}
\begin{split}
\braket{\prod_{i=1}^4\zeta_{\ell_im_i}(k_i)} &= \tau_s \int \frac{{\rm d}^3x{\rm d}^3y{\rm d}^3 K}{(2\pi)^3} \[\prod_i {\rm d}^2 \hat k_i 
\, \Ys{\ell_i}{m_i}(\vers k_i)\] e^{{\rm i} (\k_{12} -\vec K)\cdot\x + {\rm i} (\k_{34}+\vec K)\cdot\y} \ t\,^{\k_1 \k_2}_{\k_3 \k_4}(K,s) \\
&\;\;\;\;+ \text{$23$ perms.}
\end{split}
\ee
After having expanded the exponentials using the Rayleigh expansion, and performed the integrals using the orthonormality of the spherical harmonics and the Gaunt integral, we get
\begin{equation}
\label{eq:primordial_trispectrum_multipole_sp}
\begin{split}
\avg{\prod_{i=1}^4\zeta_{\ell_im_i}(k_i)} &= \sum_{L,M} (-1)^M
{
\begin{pmatrix} 
\ell_1 \! & \! \ell_2 \! & \! L \\
-m_1 \! & \! -m_2 \! & \! M 
\end{pmatrix}
\begin{pmatrix} 
\ell_3 \! & \! \ell_4 \! & \! L \\
-m_3 \! & \! -m_4 \! & \! -M 
\end{pmatrix} 
}
(2\pi)^3 \, \tau_s \ t^{k_1k_2 \, \ell_1\ell_2}_{k_3k_4 \, \ell_3\ell_4}(L,s) \\
&\;\;\;\;+ \text{$23$ perms.}
\end{split}
\end{equation}
In this expression, the matrices stand for the Wigner $3$-${\rm j}$ symbol and we redefined the function $t$ as
\begin{equation}
\label{eq:reduced_trispectrum_multipole_space}
\begin{split}
t^{k_1k_2 \, \ell_1\ell_2}_{k_3k_4 \, \ell_3\ell_4}(L,s) &= 
\frac{16 (2\pi)^2}{2s+1} \sum_{L_1,L_3,L'} (-1)^{\ell_{1234} + \frac12 (L_{13} + \ell_{13})+L'+L+s} \ \I^{\ell_1\ell_2}_{\ell_3 \ell_4} (L_1,L_3,L',s,L)\,\times \\
&\;\;\;\;\int_0^\infty{\rm d}x{\rm d}y\,(xy)^2 F_{L}({\Delta},x,y) j_{L_1}(k_1 x) j_{\ell_2}(k_2 x) j_{L_3}(k_3 y) j_{\ell_4}(k_4 y) P_\zeta(k_1)P_\zeta(k_3)\,\,.
\end{split}
\end{equation}
In the above equation, the function $F_L$ is defined as 
\be
\label{f_l_xy}
F_L(\Delta, x,y)=\frac{2}{\pi} \int_0^{\infty}{\rm d}K \, K^{2+2\Delta} \, P_\zeta(K) \, j_L(Kx) \, j_L(Ky)\,\,.
\ee
while the symbol $\I$ is 
\be
\label{ang_mod_projection}
\I ^{\ell_1 \ell_2}_{\ell_3 \ell_4}(L_1,L_3,L',s,L) = (2L+1) \ h^{L_1 \, \ell_2 \, L'} \ h^{L_3 \, \ell_4 \, L'} \ h^{\ell_1 \, L_1 \, s} \ h^{\ell_3 \, L_3 \, s} \
{
\begin{Bmatrix} 
\ell_1 \! & \! \ell_2 \! & \! L \\
L' \! & \! s \! & \! L_1 
\end{Bmatrix}
\begin{Bmatrix} 
\ell_3 \! & \! \ell_4 \! & \! L \\
L' \! & \! s \! & \! L_3 
\end{Bmatrix}\,\,.
}
\ee

\subsection{CMB \texorpdfstring{$3$- and $4$-point functions}{3- and 4-point functions}}

Given the primordial signal in multipole space, Eqs.~\eqref{primordial_bispectrum_multipole_sp},~\eqref{reduced_bispectrum_multipole_space} 
and Eqs.~\eqref{eq:primordial_trispectrum_multipole_sp},~\eqref{eq:reduced_trispectrum_multipole_space} respectively, 
we can easily obtain the CMB $3$- and $4$-point functions in harmonic space, using the relation 
\be
a_{\ell m} = 4\pi(-{\rm i})^\ell \int \frac{{\rm d}k}{(2\pi)^3}\,k^2\,\Delta_\ell(k)\,\zeta_{\ell m} (k)\,\,,
\ee
where $\Delta_\ell(k)$ is the temperature transfer function (we drop the superscript ``$T$'' for simplicity of notation). 

\subsubsection{CMB bispectrum}

The CMB bispectrum is given by
\be
\label{cmb_bisspectrum}
\begin{split}
\avg{\prod_{i=1}^3 a_{\ell_im_i}} &= \[ \prod_{i=1}^3 4\pi (-{\rm i})^{\ell_i} \int \frac{{\rm d}k_i}{(2\pi)^3}\,k_i^2\,\Delta_{\ell_i}(k_i) \] \avg{\prod_{i=1}^3 \zeta_{\ell_im_i} } \\
&=
{
\begin{pmatrix} 
\ell_1 \! & \! \ell_2 \! & \! \ell_3 \\
m_1 \! & \! m_2 \! & \! m_3 
\end{pmatrix}
}
\,\F_s \, b^{\ell_1 \, \ell_2 \, \ell_3} (s) + (\vec{k}_2 \to \vec{k}_3)\,\,.
\end{split}
\ee
The primordial information is encoded in the reduced bispectrum $b^{\ell_1 \, \ell_2 \, \ell_3}(s)$ whose expression is
\be\label{reduced_bispectrum}
b^{\ell_1 \, \ell_2 \, \ell_3} (s) = \frac{4\pi}{2s+1} \sum_{L_1 , L_2} I^{\ell_1\,\ell_2\,\ell_3}(L_1,L_2,s) \, {R_{\rm bis}}\, ^{\ell_1 \ell_2 \ell_3}_{L_1 L_2} (\Delta)\,\,.
\ee
In the above expression, the function $R_{\rm bis}$ is given by
\be
\label{bis_integral}
{R_{\rm bis}}\, ^{\ell_1 \ell_2 \ell_3}_{L_1 L_2} (\Delta) = \int_0^{\infty} {\rm d}x \,x^2 \ \beta_{\ell_1 L_1}(\Delta, x) \ \beta_{\ell_2 L_2}(-\Delta, x) \ \alpha_{\ell_3}(x)\,\,,
\ee
where $\alpha$ and $\beta$ are defined as
\begin{align}
\alpha_\ell(x) &= \frac 2\pi \int_0^\infty {\rm d}k \ k^2 \ \Delta_\ell(k) \ j_\ell(kx)\,\,, \label{alpha_ell} \\
\beta_{\ell L}(\Delta, x) &= \frac 2\pi \int_0^\infty {\rm d}k \ k^{2+\Delta} \ P_\zeta(k) \ \Delta_\ell(k) \ j_L(kx)\,\,. \label{beta_ell}
\end{align}

\subsubsection{CMB trispectrum}

The CMB trispectrum, instead, is given by
\begin{equation}
\label{cmb_trispectrum}
\begin{split}
\avg{\prod_{i=1}^4 a_{\ell_im_i}} &= \[ \prod_{i=1}^4 4\pi (-{\rm i})^{\ell_i} \int \frac{{\rm d}k_i}{(2\pi)^3}\,k_i^2\,\Delta_{\ell_i}(k_i) \] \avg{\prod_{i=1}^4 \zeta_{\ell_im_i} }\,\times \\
&\;\;\;\; \sum_{L,M} (-1)^M
{
\begin{pmatrix} 
\ell_1 \! & \! \ell_2 \! & \! L \\
m_1 \! & \! m_2 \! & \! -M 
\end{pmatrix}
\begin{pmatrix} 
\ell_3 \! & \! \ell_4 \! & \! L \\
m_3 \! & \! m_4 \! & \! M 
\end{pmatrix} 
}
\, \tau_s \ t^{\ell_1 \ell_2}_{\ell_3 \ell_4} (L,s) + \text{$23$ perms.}\,\,,
\end{split}
\end{equation}
where $t^{\ell_1 \ell_2}_{\ell_3 \ell_4} (L,s)$ is the reduced trispectrum, defined as
\be 
\label{reduced_trispectrum}
\begin{split}
t^{\ell_1 \ell_2}_{\ell_3 \ell_4} (L,s) = 
\frac{4\pi}{2s+1} &\sum_{L_1,L_3,L'} (-1)^{\frac{L_{13}}{2} + \frac{\ell_{13}}{2} + \ell_{1234}+L+L'+s} \ \I ^{\ell_1 \ell_2}_{\ell_3 \ell_4}(L_1,L_3,L',s,L)\, \times \\
& \int_0^\infty {\rm d}x {\rm d}y\,(xy)^2 \ \beta_{\ell_1 L_1}(-\Delta,x) \ \alpha_{\ell_2}(x) \ \beta_{\ell_3 L_3}(-\Delta, y) \ \alpha_{\ell_4}(y) F_L({\Delta},x,y)\,\,.
\end{split}
\ee
The function $\alpha$ is peaked at the recombination distance $r_*$. 
Then, if $F_L({\Delta},x,y)$ varies slowly around that point, the integrals in $x$ and $y$ become separable, 
and we can approximate the function $F_L({\Delta},x,y)$ as $F_L({\Delta},r_*,r_*)$. 
This greatly reduces the computational cost needed to estimate the signal. 
In Appendix \ref{appendix-B} we show that this assumption is satisfied for small $L$, that is where most of the signal for the trispectrum is. 
The second line of Eq.~\eqref{reduced_trispectrum} can be \mbox{approximated to}
\be
R_{\rm tris}^{\ell_1 L_1\ell_2}(\Delta) \ R_{\rm tris}^{\ell_3 L_3 \ell_4}(\Delta) \ F_L({\Delta},x,y)\,\,,
\ee
where $R_{\rm tris}$ is
\be\label{tris_integral}
R_{\rm tris}^{\ell_1 L_1\ell_2}(\Delta) = \int_0^\infty {\rm d}x \, x^2 \, \beta_{\ell_1L_1}(-\Delta, x) \, \alpha_{\ell_2}(x)\,\,.
\ee

\begin{center}
***
\end{center}

Before concluding this section let us study the signal for $s=0$.\footnote{In this case one recovers the standard parameterizations: $\F_0 = \fnl$ and $\tau_0 = \tau_{\rm NL}$.} 
In this situation the primordial correlators are enhanced thanks to the exchange of a long-wavelength scalar 
field and we recover the known results of, \eg, Refs.~\cite{Liguori:2005rj,Kogo:2006kh}. 
Thanks to the properties of the Wigner $6$-${\rm j}$ symbols, the functions $I$ and $\I$ simplify greatly if $s=0$: they reduce to 
\begin{align}
&I ^{\ell_1 \, \ell_2 \, \ell_3}(L_1,L_2,0) = \bigg(\frac{h^{\ell_1\,\ell_2\,\ell_3}}{4\pi}\bigg)\, \delta_{\ell_1 L_1} \delta_{\ell_2 L_2}\,\,, \\
&\I ^{\ell_1 \ell_2}_{\ell_3 \ell_4}(L_1,L_3,L',0,L) = \bigg(\frac{h^{\ell_1\,\ell_2\,L} \ h^{\ell_3\,\ell_4\,L}}{4\pi}\bigg)\, \delta_{\ell_1 L_1} \delta_{\ell_3 L_3}\delta_{L L'}\,\,.
\end{align}
The final expressions for the CMB $3$- and $4$-point functions, then, are
\be
\label{bispectrum_CMB_signal_s=0}
b^{\ell_1 \, \ell_2 \, \ell_3} (0) =
{
\begin{pmatrix} 
\ell_1 \! & \! \ell_2 \! & \! \ell_3 \\
-m_1 \! & \! -m_2 \! & \! m_3 
\end{pmatrix}
}
\,{\F_0} \ h^{\ell_1\,\ell_2\,\ell_3} \ {R_{\rm bis}}\, ^{\ell_1 \ell_2 \ell_3}_{\ell_1 \ell_2} (\Delta)\,\,, 
\ee
and 
\begin{equation}
\label{eq:trispectrum_CMB_signal_s=0}
\begin{split}
t^{\ell_1 \ell_2}_{\ell_3 \ell_4} (L,0) 
= \sum_{L, M}\,&(-1)^M
{
\begin{pmatrix} 
\ell_1 \! & \! \ell_2 \! & \! L \\
-m_1 \! & \! -m_2 \! & \! M 
\end{pmatrix}
\begin{pmatrix} 
\ell_3 \! & \! \ell_4 \! & \! L \\
-m_3 \! & \! -m_4 \! & \! -M 
\end{pmatrix} 
}
\, \times \\
&\,{\tau_0} \ h^{\ell_1\,\ell_2\,L} \ h^{\ell_3\,\ell_4\,L} \ R_{\rm tris}^{\ell_1 \ell_1\ell_2}(\Delta) \ R_{\rm tris}^{\ell_3 \ell_3 \ell_4}(\Delta) \ F_L(r_*,r_*)\,\,.
\end{split}
\end{equation}
If the field is massless, i.e.~$\Delta=0$ (or equivalently $\nu=3/2$), the above expressions agree with already known results \cite{Liguori:2005rj,Kogo:2006kh}.

\section{Analysis}
\label{analysis_sec}

We are now in position to estimate the constraints we can get on the coefficients $\F_s$ and $\tau_s$ for different values of the spin and the mass of the exchanged particle. 
We consider a noise-free and cosmic-variance-limited experiment measuring temperature anisotropies up to a maximum multipole of $\ell_{\rm max}$. 
We also assume that the non-Gaussian signal is very weak, so that we can neglect the non-Gaussian contribution to the cosmic variance. 
Let us define the Fisher matrix $F$ (which for us is a $1\times 1$ matrix) for the bispectrum as \cite{Shiraishi:2013vja} 
\be
\label{StN_bis}
F_{\rm bis} = \sum_{\ell_1>\ell_2>\ell_3 \geq 4} \frac{\abs{B_{\ell_1 \ell_2 \ell_3} (s)}^2}{C_{\ell_1}C_{\ell_2}C_{\ell_3}}\,\,,
\ee
where $B_{\ell_1 \ell_2 \ell_3} (s)$ is given by all the possible permutations of Eq.~\eqref{reduced_bispectrum}. 
The Fisher matrix for the trispectrum is instead defined as \cite{Hu:2001fa}
\be\label{StN_tris}
F_{\rm tris} = \sum_L \sum_{\ell_1>\ell_2>\ell_3>\ell_4\geq 5} \frac{|T^{\ell_1 \ell_2}_{\ell_3 \ell_4} (L,s)|^2}{(2L+1)C_{\ell_1}C_{\ell_2}C_{\ell_3}C_{\ell_4}}\,\,,
\ee
where $T^{\ell_1 \ell_2}_{\ell_3 \ell_4}$ is the trispectrum averaged over possible orientations of quadrilaterals, \ie~
\be
\begin{split}
T^{\ell_1 \ell_2}_{\ell_3 \ell_4} (L,s) = P^{\ell_1 \ell_2}_{\ell_3 \ell_4} (L,s) + (2L+1)\sum_{L'} \bigg[ 
&(-1)^{\ell_{23} }
{
\begin{Bmatrix} 
\ell_1 \! & \! \ell_2 \! & \! L \\
\ell_4 \! & \! \ell_3 \! & \! L' 
\end{Bmatrix}
}
\,P^{\ell_1 \ell_3}_{\ell_2 \ell_4} (L',s) \\
&+ (-1)^{L+L'} 
{
\begin{Bmatrix} 
\ell_1 \! & \! \ell_2 \! & \! L \\
\ell_3 \! & \! \ell_4 \! & \! L' 
\end{Bmatrix}
}
\,P^{\ell_1 \ell_4}_{\ell_3 \ell_2} (L',s)\bigg]\,\,, 
\end{split}
\ee
and $P^{\ell_1 \ell_4}_{\ell_3 \ell_2} $ is defined as
\be
P^{\ell_1 \ell_4}_{\ell_3 \ell_2} (L,s) = 2 \[ t^{\ell_1 \ell_2}_{\ell_3 \ell_4}(L,s) +(-1)^{\ell_{12}+L} \, t^{\ell_2 \ell_1}_{\ell_3 \ell_4} (L,s)+(-1)^{\ell_{34}+L} \, t^{\ell_1 \ell_2}_{\ell_4 \ell_3} (L,s)+(-1)^{\ell_{1234}} \, t^{\ell_2 \ell_1}_{\ell_4 \ell_3} (L,s) \]\,\,.
\ee

\subsection{Simple estimates}

Before proceeding with the full analysis, let us first estimate the expected behavior of both {$F_{\rm bis}$ and $F_{\rm tris}$} as a function of the maximum multipole $\ell_{\rm max}$. 
As we are going to confirm in the next section, the value of $s$ does not affect the scaling of the signal-to-noise ratio with $\ell_{\rm max}$, 
so we fix $s=0$ for simplicity (correspondingly, we can fix $L=\ell$ in the $\beta$ functions of Eq.~\eqref{beta_ell}). 
From Eq.~\eqref{bispectrum_CMB_signal_s=0}, we see that the estimating ${b^{\ell_1\,\ell_2\,\ell_3}}$ requires an estimate 
of ${R_{\rm bis}}\, ^{\ell_1 \ell_2 \ell_3}_{\ell_1 \ell_2} (\Delta)$. To do this we approximate the radiation transfer function as a spherical Bessel function, 
neglecting acoustic physics, \ie~we use the Sachs-Wolfe approximation (see Appendix \ref{appendix-C} for more details). In this approximation we have 
that $\Delta_\ell(k) = - j_\ell(k r_*)/5$, where $r_*$ is the comoving distance to the last-scattering surface. Consequently, Eq.~\eqref{alpha_ell} becomes simply 
$\alpha_\ell(x) = {-{\delta(r_\ast-x)}/{5 \sqrt{r_*^3 \, x}}}\,\,\forall\,\ell$, and from Eq.~\eqref{bis_integral} 
we get ${R_{\rm bis}}\, ^{\ell_1 \ell_2 \ell_3}_{\ell_1 \ell_2} (\Delta)\sim \beta_{\ell_1{\ell_1}}(\Delta) \beta_{\ell_2{\ell_2}}(-\Delta)$. 
From Eq.~\eqref{beta_ell} we finally get $\beta_{\ell{\ell}} (\Delta) \propto \ell^\Delta C_\ell$. 
Then, a {na{\" i}ve} estimate of the signal-to-noise ratio for the bispectrum is (see also \cite{Babich:2004yc})
\be
\label{bis_estimate}
F_{\rm bis} \approx \int{\rm d}^2\ell_1{\rm d}^2\ell_2 \ \frac{{(b^{\ell_1\,\ell_2\,\ell_3})^2}}{C_{\ell_1} C_{\ell_2}^2} \propto \ell_{\rm max}^2\,\,.
\ee 
Interestingly, the scaling of $F_{\rm bis}$ with $\ell_{\rm max}$ is independent of the mass of the exchanged particle.\footnote{Strictly speaking, 
if the exchanged particle is massless the dependence of {$F_{\rm bis}$} on $\ell_{\rm max}$ is slightly different: 
$F_{\rm bis}\propto \ell_{\rm max}^2 \log \ell_{\rm max}$. 
The reason for this difference lies in the fact that for $\Delta \neq 0$ our template explicitly chooses $\vec{k}_1$ as the long mode, 
so that we can integrate Eq.~\eqref{bis_estimate} only in the region $\ell_1\leq \ell_2$. For $\Delta = 0$, instead, we can consider the contribution 
coming from all the configurations of $\ell_1$ and $\ell_2$. }
Something similar happens also for the trispectrum. 
It is easy to see that, in the counter-collapsed configuration, $t_{\ell_3 \ell_4}^{\ell_1 \ell_2} (L,s) \approx F_{L}(\Delta) \beta_{\ell_1 \ell_1} (\Delta) \beta_{\ell_3 \ell_3} (\Delta) $, 
with $F_{L}(\Delta) \propto L^{2\Delta} C_L$. This leads to the estimate\footnote{Notice that the trispectrum template 
is valid in any configuration of the momenta and therefore we are not assuming any hierarchy between the multipoles $\ell_1$ and $\ell_2$.} 
\begin{equation}
\label{tris_estimate}
F_{\rm tris} \approx \int{\rm d}^2L {\rm d}^2 \ell_1{\rm d}^2 \ell_2 \, \frac{\big(t^{\ell_1 \ell_2}_{\ell_3 \ell_4} (L)\big)^2}{C_{\ell_1}^2 C_{\ell_2}^2} \propto
\begin{cases}
\ell_{\rm max}^{4(1-\Delta)} & \text{if $0\leq\Delta<1/2\,\,,$} \\ 
\ell_{\rm max}^2 & \text{if $1/2<\Delta<1\,\,.$}
\end{cases}
\end{equation}
For $\Delta = 1/2$ and $\Delta = 1$ the scaling is slightly different: $F_{\rm tris} \propto \ell_{\rm max}^2 \log\ell_{\rm max}$ 
and $F_{\rm tris} \propto\ell_{\rm max}^2 \log^2\ell_{\rm max}$, respectively.

\subsection{Numerical analysis}

The outcome of these estimates motivates us to study the behavior of the signal-to-noise for different values of $\Delta$ 
since we see that, at least in the bispectrum case, the scaling with $\ell_{\rm max}$ does not drop off 
as we increase the mass of the exchanged particle. We perform the forecasts for $s=0,2,4$ and $\Delta=0, 1/2$ and $1$. 
The choices $\Delta = 0, 1$ are particularly interesting since the former corresponds to the exchange of a massless particle, 
while the latter corresponds to the Higuchi Bound for particles with $s \geq 1$.
It is therefore the smallest value of $\Delta$ (\ie~the strongest scale dependence of the bispectrum in the squeezed limit, 
$\avg{\zeta_{\k_1} \zeta_{\k_2} \zeta_{\k_3}}'\sim k_1^{\Delta-3}$ for $k_1\to 0$) 
one can get if the spinning particle does not couple with the foliation provided by the inflaton.

\begin{figure}[H]
\centering
\begin{tabular}{l}
\includegraphics[width=0.475\columnwidth]{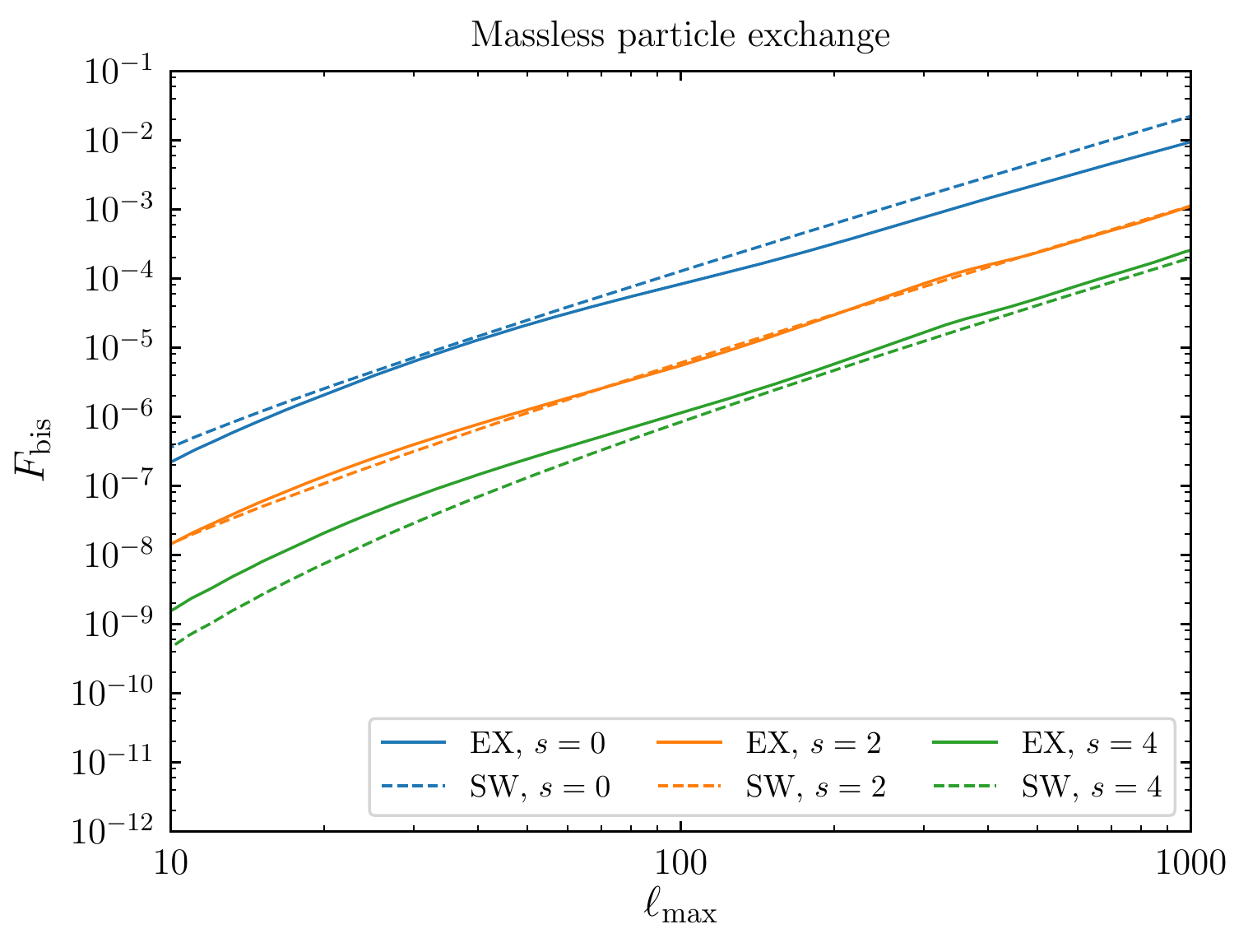} 
\includegraphics[width=0.475\columnwidth]{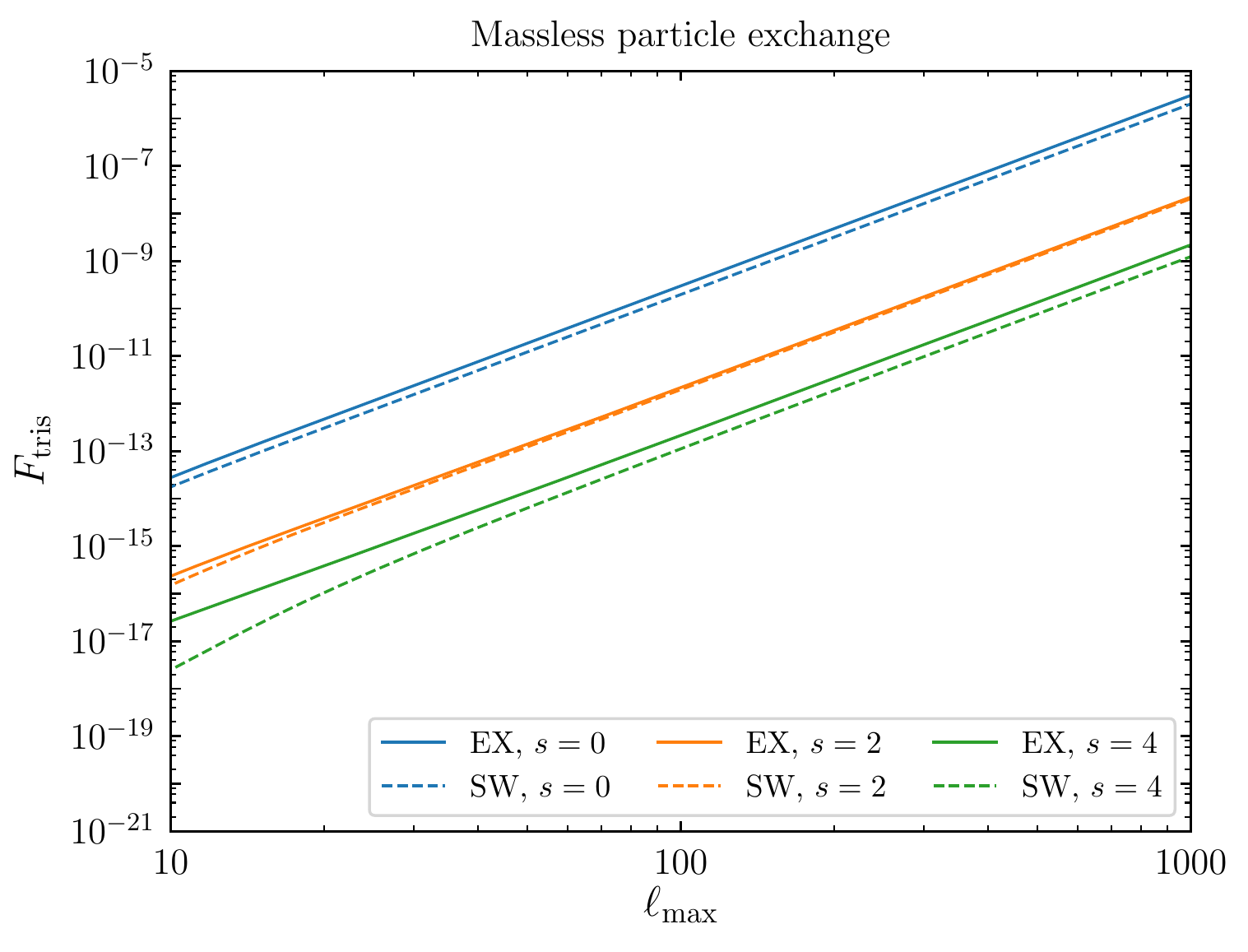}
\end{tabular}

\begin{tabular}{l}
\includegraphics[width=0.475\columnwidth]{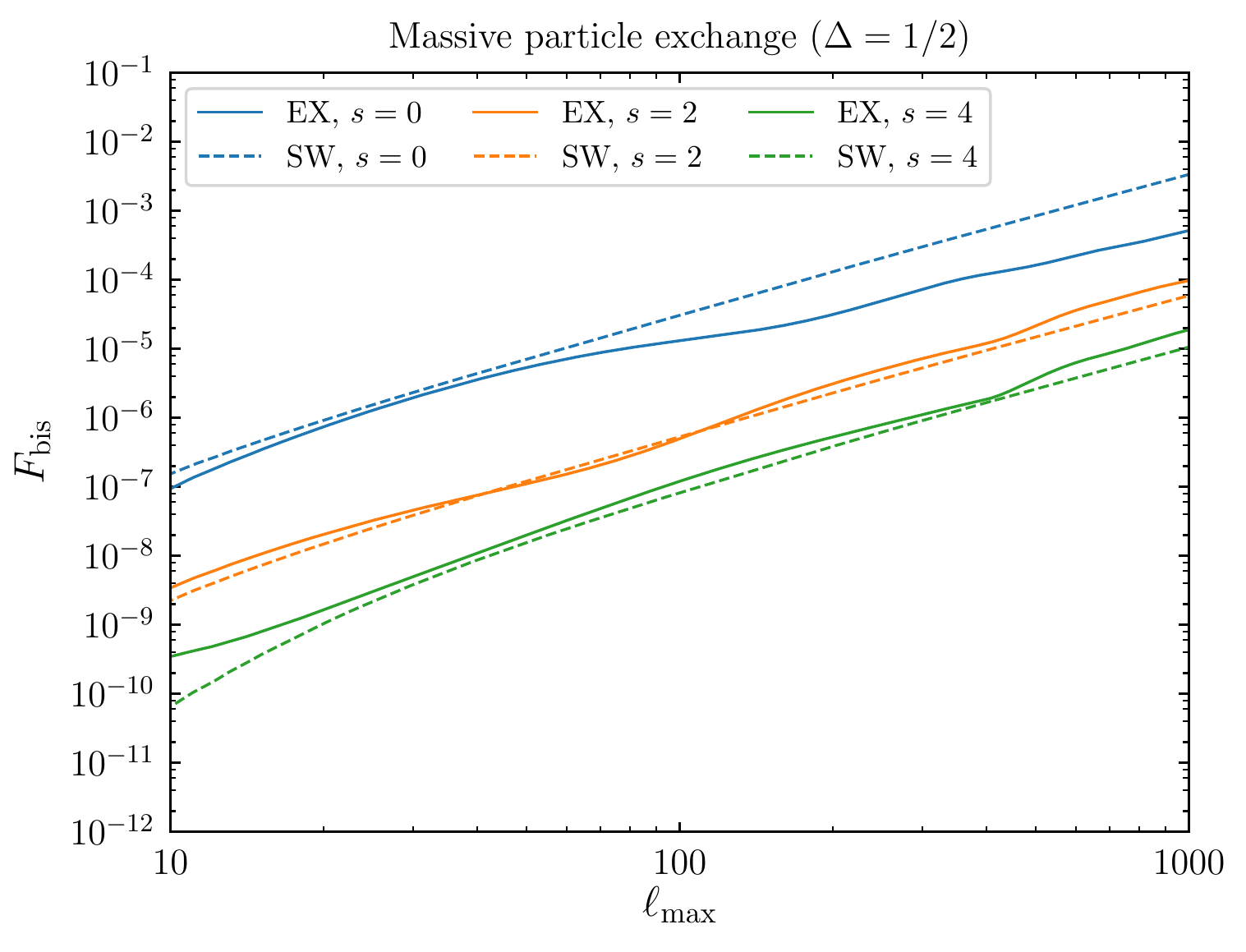} 
\includegraphics[width=0.475\columnwidth]{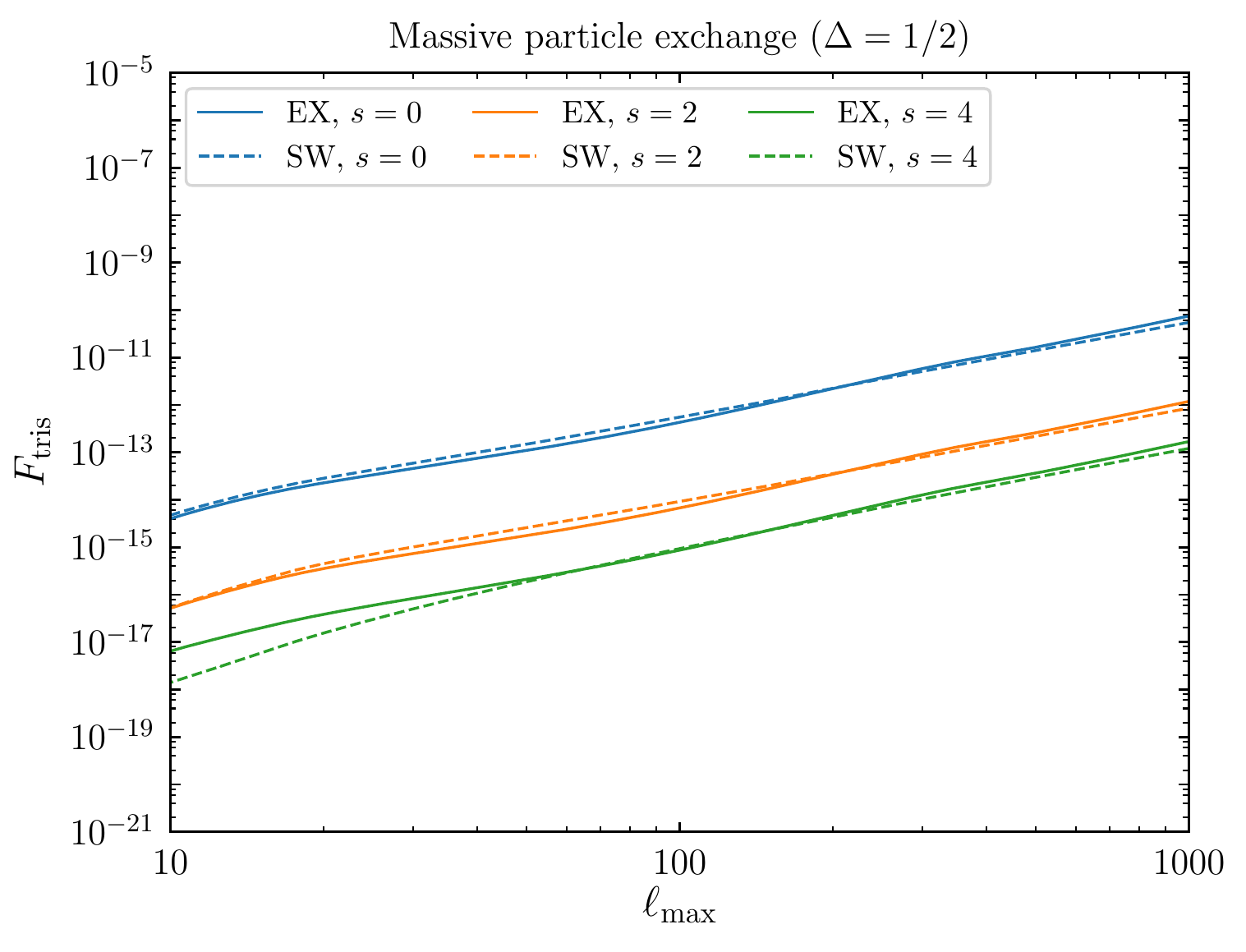}
\end{tabular}

\begin{tabular}{c}
\includegraphics[width=0.475\columnwidth]{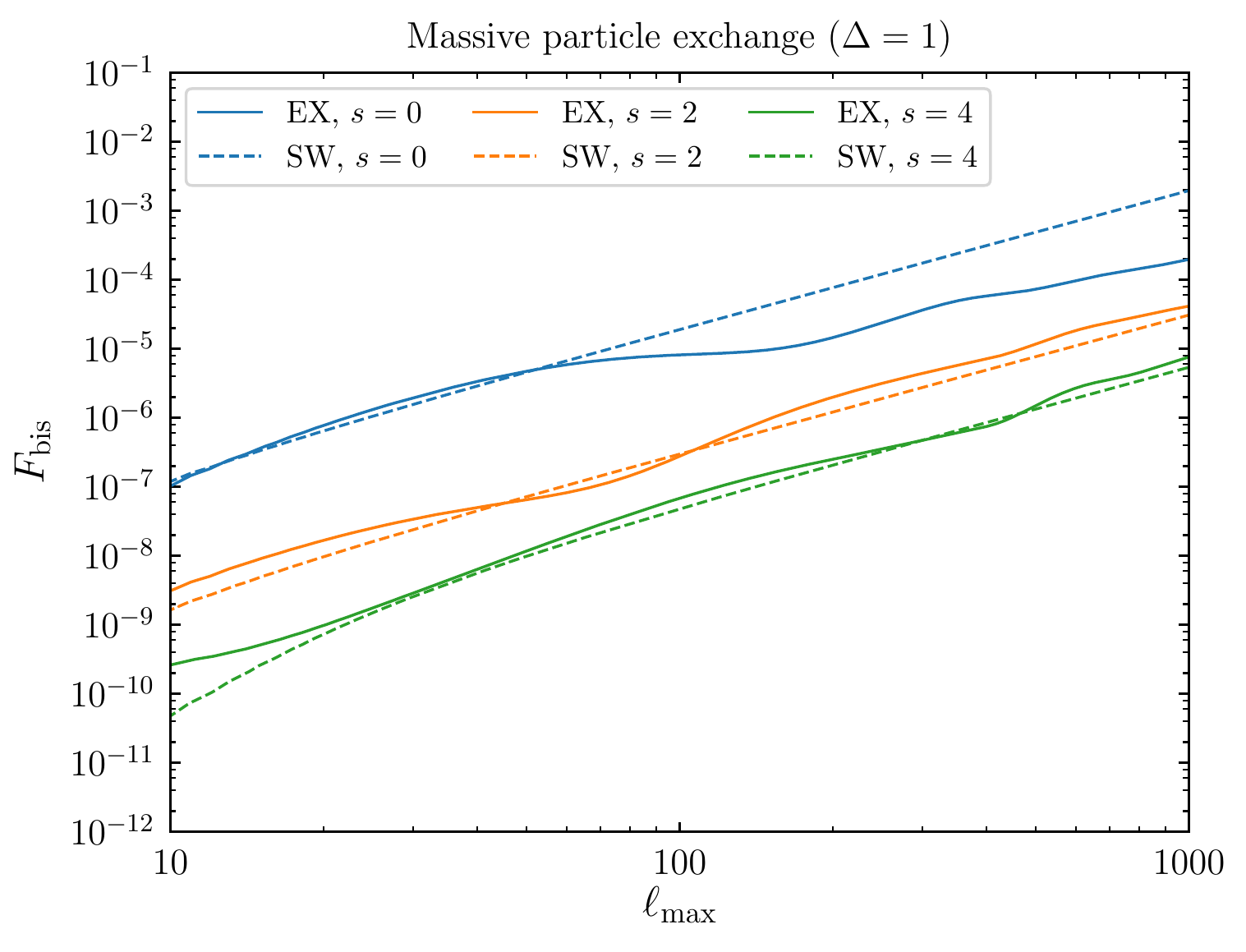} 
\end{tabular}
\caption{From top to bottom: Fisher matrix for the bispectrum (left panels) and trispectrum (right panels) arising by the exchange of a massless particle with $s=0,2,4$, 
and with $\Delta=0,1/2$. The solid lines make use of the full radiation transfer function, while the dashed lines show the SW approximation. 
{$F_{\rm tris}$} has been computed summing only over soft-limit contributions, $L\leq 10$. 
The bottom panel shows the Fisher matrix for the bispectrum for $\Delta=1$, and $s=0,2,4$. In all plots ${\cal F}_s$ and $\tau_s$ are fixed to $1$.} 
\label{fig_signal_to_noises}
\end{figure}

The plots in the top panels of Fig.~\ref{fig_signal_to_noises} show the behavior of the signal-to-noise for the bispectrum and the trispectrum given by the exchange of a massless particle 
as a function of $\ell_{\rm max}$.\footnote{In the plots we fix ${\cal F}_s$ and $\tau_s$ to $1$. 
Note that in the event of a positive detection, our estimates of the signal-to-noise ratio break down due to the non-Gaussian contribution to the noise. 
Beyond this point, an improved estimator is necessary to decrease the error bars as $1/\sqrt{N_{\rm data}}$.} 
Since the trispectrum template is peaked in the counter-collapsed configuration, we have truncated the sum on $L$, considering only the contributions with $L\leq 10$. 
This simplification dramatically reduces the computational time required to perform the evaluation of the signal-to-noise, 
and allows us to perform the analysis in a reasonable amount of time. 
We confirm the scaling behavior predicted in Eqs.~\eqref{bis_estimate},~\eqref{tris_estimate}: as expected it is not affected by the value of the particle spin. 
As we vary $s$, what changes is the amplitude of the signal, that drops as the spin increases. 
The numerical results are approximately fitted by 
\begin{align}
F_{\rm bis} & \approx \frac{\num{6.95d-10}}{s^2+0.50} \ \F_s^2 \ \ell_{\rm max}^2\, \log\ell_{\rm max}\,\,, \label{fit_bis_massless} \\
F_{\rm tris} & \approx \frac{\num{3.60d-19}}{s^4+0.12} \ \tau_s^2 \ \ell_{\rm max}^4\,\,. \label{fit_tris_massless}
\end{align}

{The Fisher matrixes in the massive case are also shown in Fig.~\ref{fig_signal_to_noises} for both $\Delta = 1/2$ and $\Delta=1$.} 
Again, we confirm the predicted scalings, ($F_{\rm bis} \propto \ell^{2}_{\rm max}$ and $F_{\rm tris} \propto \ell_{\rm max}^2\log\ell_{\rm max}$) 
with the possible exception of the bispectrum in the $s=0$ case (that deviates from the predicted scaling for $\ell_{\rm max} \gtrsim 400$). 
Even in the massless case the signal-to-noise for $s=0$ presents such a deviation, as we see from 
the top left panel of Fig.~\ref{fig_signal_to_noises} (we also notice that this behavior at large $\ell_{\rm max}$ 
is consistent with the one found by Ref.~\cite{Franciolini:2018eno}: see the blue line in their Fig.~4). However this unexpected behavior becomes more pronounced in the massive case. 
It is possible that this behavior is just a very slow oscillation in $\ell_{\rm max}$ around $F_{\rm bis} \propto \ell^{2}_{\rm max}$. 
In this case, more multipoles ($\ell_{\rm max}>1000$) are needed to recover the expected scaling: we leave this to a future analysis. 

{Notice that in the massive case, even if the signal is still peaked in the counter-collapsed configuration, it increases much more slowly in this limit. 
This means that, while for $\Delta = 1/2$ we only expect order one corrections to our result coming from the terms with $L>10$, we }
could not perform the analysis for the trispectrum in the $\Delta=1$ case. 
One should sum over all the possible values of $L$ to get the correct result, making the analysis practically unfeasible.\footnote{While 
the total computational time scales as $\ell_{\rm max}^2$ if the signal is peaked in the counter-collapsed configuration, 
in the most general configuration it scales as $\ell_{\rm max}^5$.} 

\begin{figure}[H]
\centering
\begin{tabular}{c}
\includegraphics[width=0.765\columnwidth]{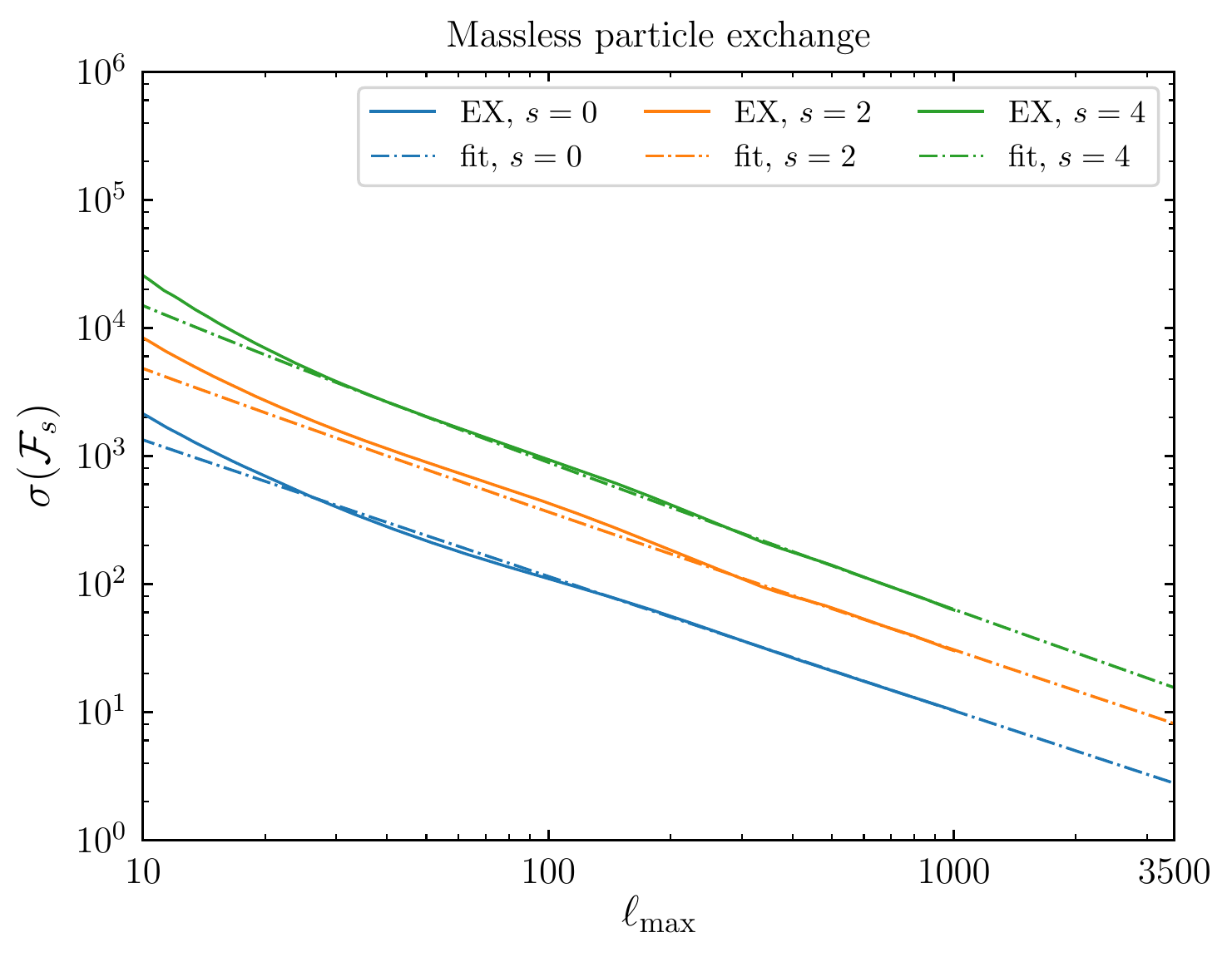} \\
\includegraphics[width=0.765\columnwidth]{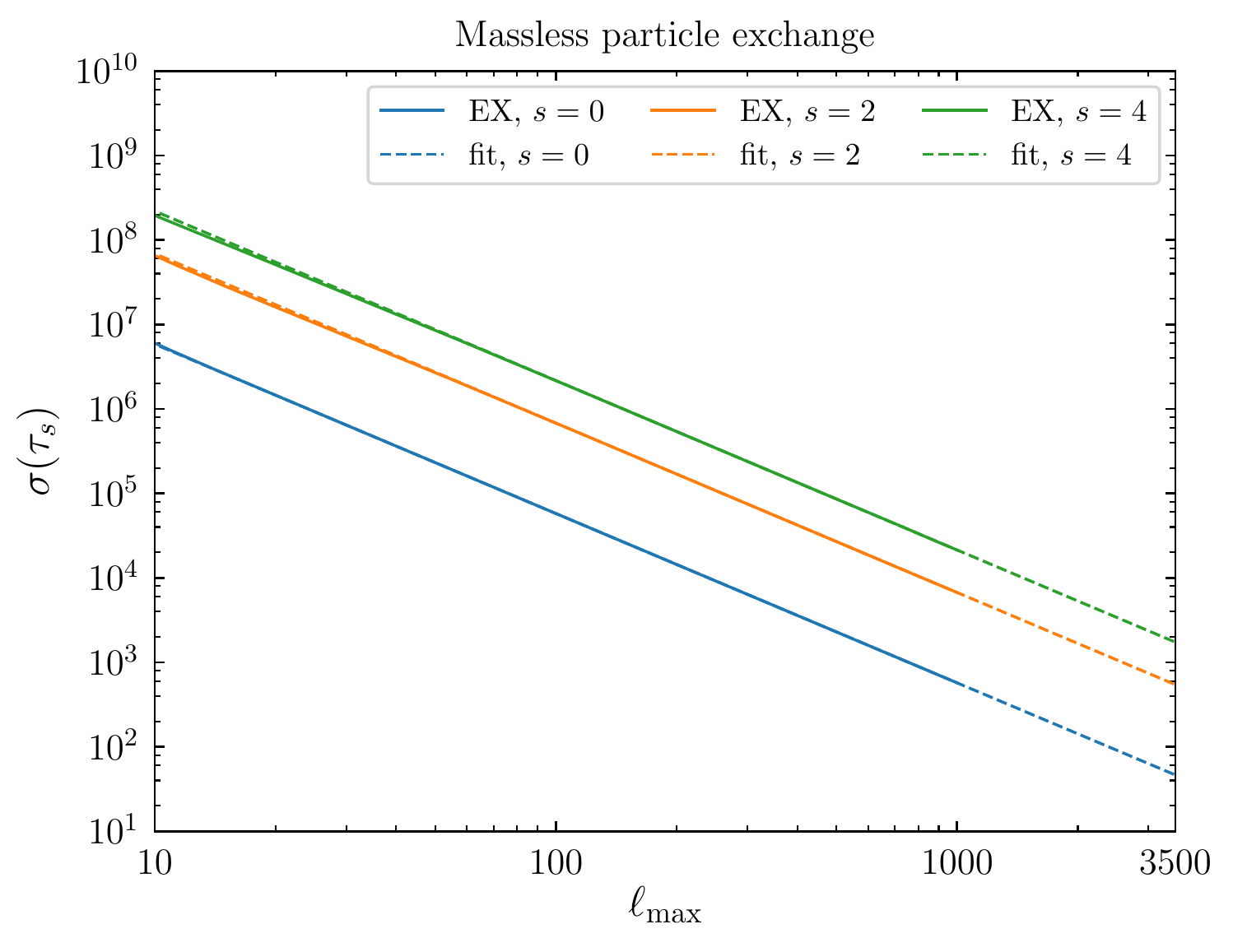}
\end{tabular}
\caption{Expected $1\sigma$ errors on $\F_s$ (top panel) and $\tau_s$ (bottom panel) for $s=0,2,4$ and $\Delta=0$ as a function of $\ell_{\rm max}$. 
The solid lines are computed using the exact expressions for the bispectrum and trispectrum up to $\ell_{\rm max}=1000$, 
while the dotted-dashed lines are their \mbox{extrapolations up to $\ell_{\rm max} = 3500$.}} 
\label{massless_errorbars}
\end{figure}

Given the results of Eqs.~\eqref{fit_bis_massless},~\eqref{fit_tris_massless} and of Fig.~\ref{fig_signal_to_noises} 
we can estimate the bounds that a cosmic-variance-limited CMB experiment could put on the amplitude of these new shapes of non-Gaussianity. 
The expected $1\s$ errors on $\F_s$ and $\tau_s$ are given by $\s({\F_s}) = 1/\sqrt{F_{\rm bis}}$ and $\s({\tau_s}) = 1/\sqrt{F_{\rm tris}}$\,. 
Fig.~\ref{massless_errorbars} shows these $1\s$ errors for $\F_s$ and $\tau_s$ for the case of a massless particle exchange extrapolated up to $\ell_{\rm max} = 3500$ 
(above this threshold lensing effects become important and we can no longer trust the power-law behavior of the signal-to-noise \cite{Komatsu:2001rj}).\footnote{We 
also stress that the effects of Silk damping start to become relevant at $\ell\gtrsim 1000$ \cite{Babich:2004yc}. 
These effects should be taken into account when extrapolating our results up to $\ell_{\rm max} = 3500$: 
however, they are expected to give a correction that scales only logarithmically with $\ell_{\rm max}$ \cite{Babich:2004yc}, 
so that the error that we are making is negligible.} 
The forecasts for $\F_0$ and $\F_2$ and $\tau_0$ are consistent with the bounds obtained by \emph{Planck} \cite{Ade:2015ava}.

\begin{figure}[H]
\centering
\begin{tabular}{c}
\includegraphics[width=0.765\columnwidth]{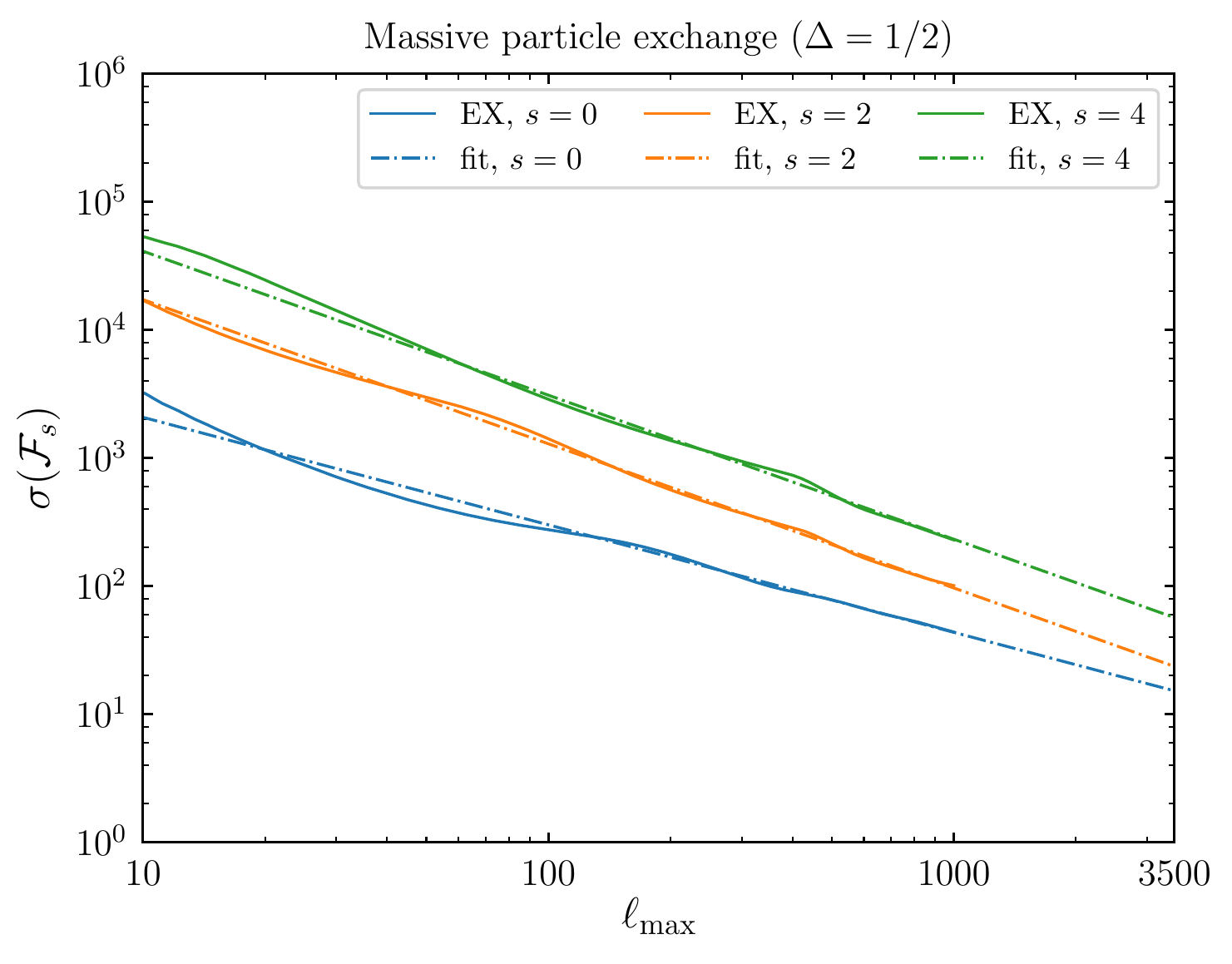} \\
\includegraphics[width=0.765\columnwidth]{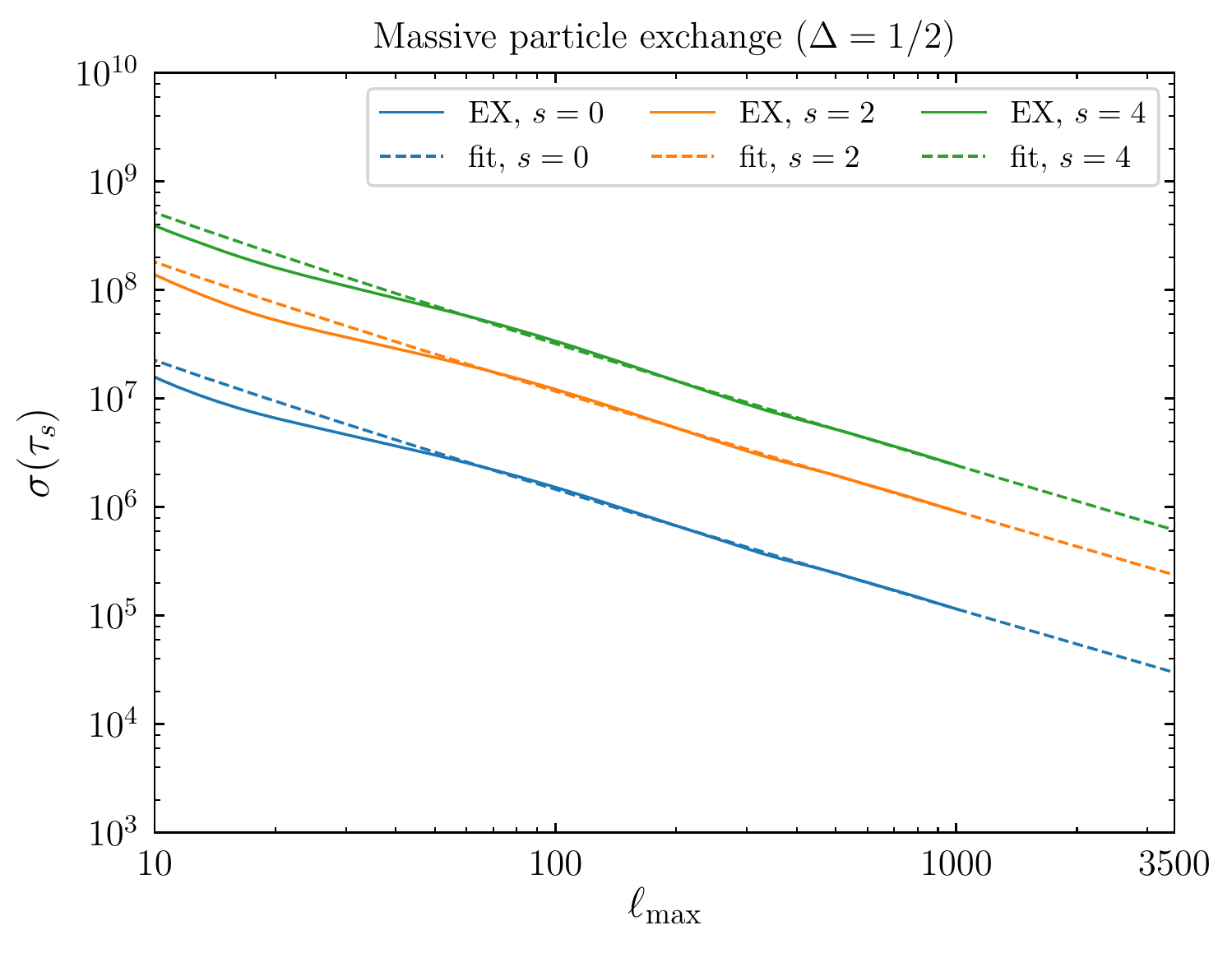}
\end{tabular}
\caption{Same as Fig~\ref{massless_errorbars}, but with $\Delta=1/2$.}
\label{Delta_05_errorbars}
\end{figure}

\begin{figure}[H]
\centering
\begin{tabular}{c}
\includegraphics[width=0.765\columnwidth]{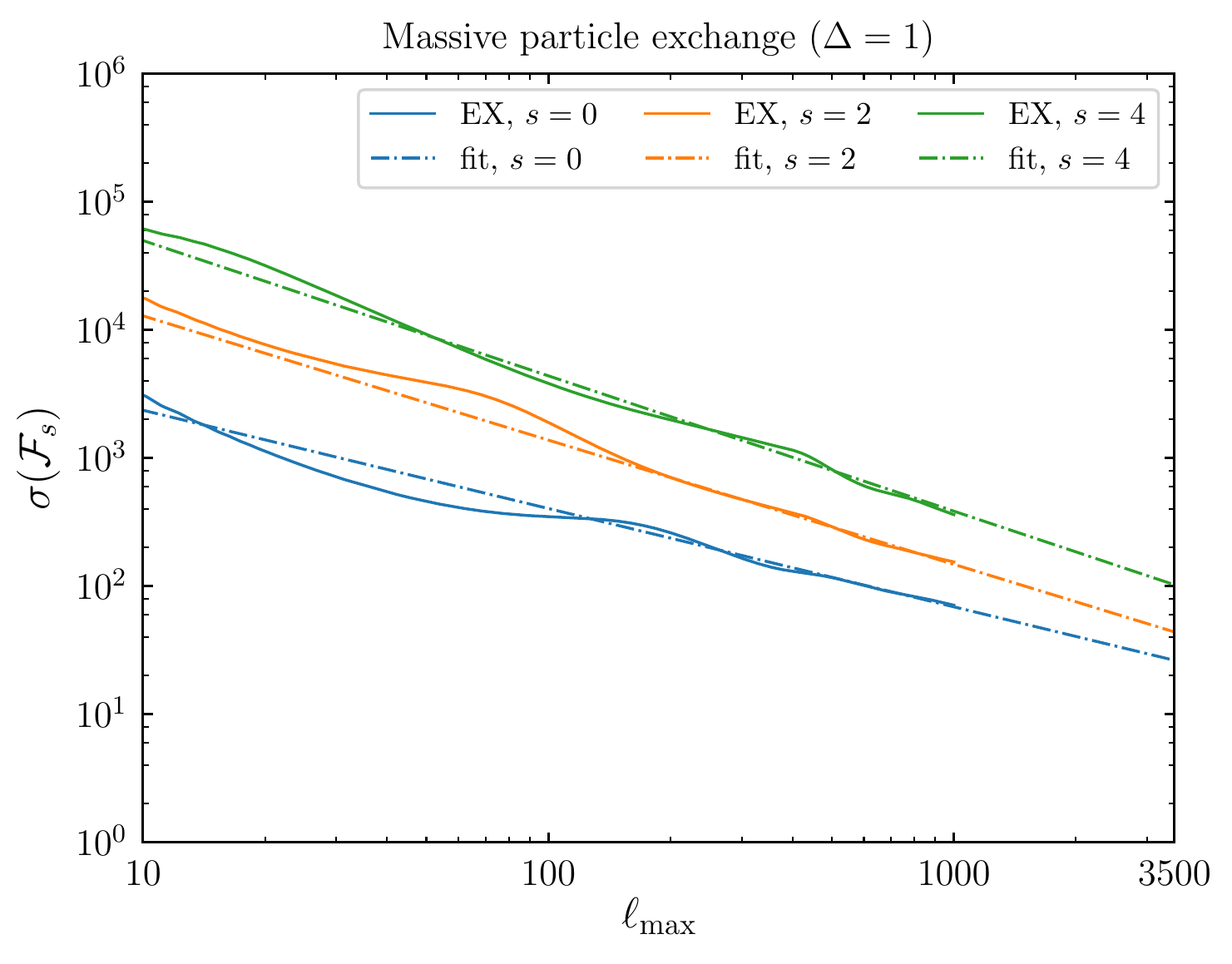}
\end{tabular}
\caption{Expected $1\sigma$ errors on $\F_s$ up to $\ell_{\rm max}=1000$, and their extrapolation up to $\ell_{\rm max} = 3500$.}
\label{massive_errorbars}
\end{figure}

Let us move now to the forecasts for the bispectrum and trispectrum amplitudes given by the exchange of a massive particle. 
{The top and bottom panels of Fig.~\ref{Delta_05_errorbars} show the expected $1\sigma$ errors on the amplitude of, respectively, the bispectrum and trispectrum 
given by the exchange of a spinning particle with $\Delta = 1/2$, while Fig.~\ref{massive_errorbars} shows the error bars for the bispectrum with $\Delta=1$.} 
{We notice that the expected error bars are of the same order of magnitude in both the massive cases considered, and are at least} 
an order of magnitude worse than those for the massless case. For instance, {for $\Delta = 1$ and $s=0$} 
we see that we could get at most $\F_0 \lesssim\O(100)$ at $\ell_{\rm max}=1000$. Moreover, following \cite{Babich:2004gb}, 
we have computed the overlap between the template of Eq.~\eqref{bis_temp} with $\Delta=1$ and the standard templates of non-Gaussianity (local, orthogonal, equilateral). 
We find that, independently of the value of the spin, the overlap between this template and the equilateral one is always quite large (the cosine being 
$\sim\text{$0.8$ -- $0.9$}$: see Appendix \ref{appendix-D}). This tells us that a dedicated analysis of these shapes 
would probably not yield better constraints on their amplitudes than the ones we can get on equilateral non-Gaussianity.

\section{Discussion and conclusions}
\label{discussion_sec}

The analysis we carried out shows that the CMB {temperature} anisotropies can still play a very important role in distinguishing between different inflationary models. 
While the \emph{Planck} Collaboration has already put bounds on some of the shapes of non-Gaussianity that we have investigated in this work, 
there are still many other shapes that are waiting to be constrained. 
For example, we stress the importance of testing the templates for the scalar trispectrum that arise in the presence of a massless spinning particle during inflation. 

We emphasize, indeed, that it is possible to have a sizable trispectrum even if the bispectrum is small. 
This happens, for instance, in models with additional higher-spin fields that do not directly mix with the inflaton field. 
The example of Section~\ref{theory_sec} shows that $\tau_s$ could be very large in these models, since it is proportional to the inverse 
of the speed of propagation of the helicity-$s$ field cubed, and such speed could be much smaller than unity. 
Our analysis shows that already with \emph{Planck} data (for which we take $\ell_{\rm max}=2500$) one could constrain 
$\tau_2 \lesssim\num{d3}$ and $\tau_4 \lesssim \num{3d3}$, 
while a future CMB experiment that will measure the temperature anisotropies up to $\ell_{\rm max} = 3500$, 
like the proposed CMB-S4 \cite{Abazajian:2016yjj}, could arrive at $\tau_2 \lesssim\num{5d2}$ and $\tau_4\lesssim\num{2d3}$. 
Given that at the moment we do not have observational constraints on the amplitude of these shapes, 
and the theoretical upper limit can be as large as {$\tau_s \lesssim\num{d5}$} (see the discussion at the end of Section \ref{theory_sec}), 
it is surely worth to look for these shapes already in both the currently available and the future CMB data. 

In this work we have also studied, for the first time in the context of CMB statistics, 
the signal-to-noise for some templates of non-Gaussianity that take into account the scale dependence due to the exchange of a massive particle with even spin.
As an example, we have studied the bispectrum {and trispectrum templates for $\Delta =1/2,1$ 
(we recall that $\Delta = 1$ corresponds} to the exchange of a particle {at the Higuchi Bound}), for different values of the spin. 
The outcome of our analysis is that, even though the signal-to-noise ratio {for the bispectrum} scales with $\ell_{\rm max}$ in the same way as in the massless case, 
its overall amplitude is {smaller}, leading, for $\Delta=1$, to ${\cal F}_0\lesssim{\cal O}(100)$ at $\ell_{\rm max}=1000$ at most. 
Moreover, this template has a large overlap with some of the standard templates already constrained by \emph{Planck}. 
This suggests that, even in the case of a detection of some level of non-Gaussianity in CMB data, 
the CMB alone cannot be used to infer the mass of the particles which were active during inflation, 
but it would need to be complemented by, for example, LSS observables: 
these could be the scale dependence of the galaxy bias, together with a modification of the bias expansion of galaxy shapes 
(given by the peculiar angular dependence of the primordial bispectrum in the squeezed limit).\footnote{See \cite{Kogai:2018nse} for a forecast using galaxy intrinsic alignments.}
We leave such analysis for future work. 

Before concluding, let us stress that in this work we have focused on the signal coming from the temperature anisotropies only. 
However, also the correlation functions involving polarization $E$-modes will add to the total signal-to-noise ratio.
In addition, a detection of $B$-modes by an experiment like CMB-S4 would also bring an extraordinary chance to test the presence of higher-spin fields.
Indeed, higher-spin fields could have a large coupling with the gravitational sector as well \cite{Bordin:2018pca} thus enhancing the correlators which involve $B$-mode polarization.

\section*{Acknowledgements}
It is a pleasure to thank Adri Duivenvoorden and Eiichiro Komatsu for very useful discussions. 
We also thank Adri Duivenvoorden, Eiichiro Komatsu and Fabian Schmidt for very useful comments on the draft. 
L.~B.~is supported by STFC Consolidated Grant No.~ST/P000703/1.
G.~C.~acknowledges support from the Starting Grant (ERC-2015-STG 678652) ``GrInflaGal'' from the European Research Council. 

\appendix

\section{Bispectrum template for exchange of a massive particle}
\label{appendix-A}

\begin{figure}[h]
\centering
\begin{tabular}{c c}
\includegraphics[width=0.475\columnwidth]{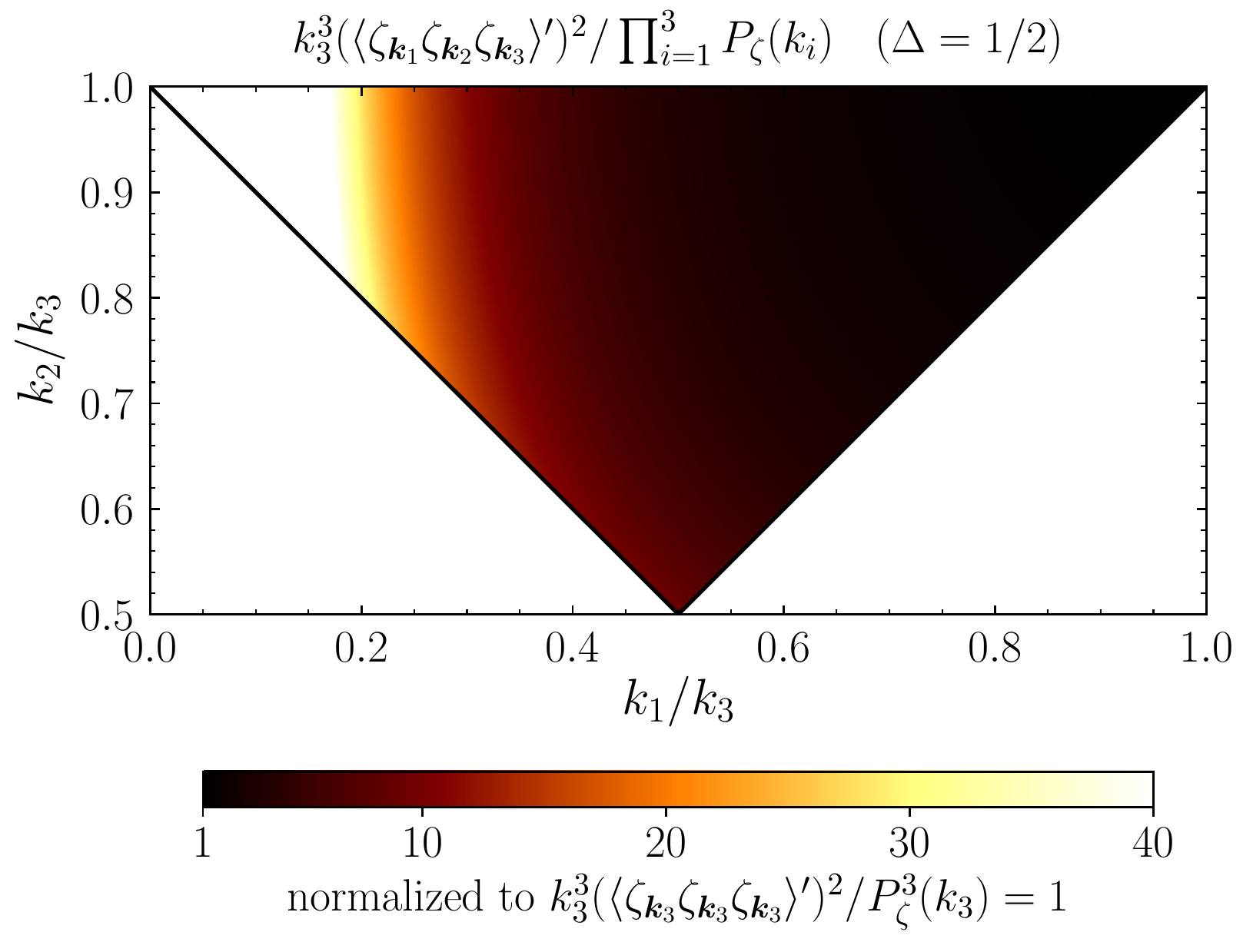} 
\includegraphics[width=0.475\columnwidth]{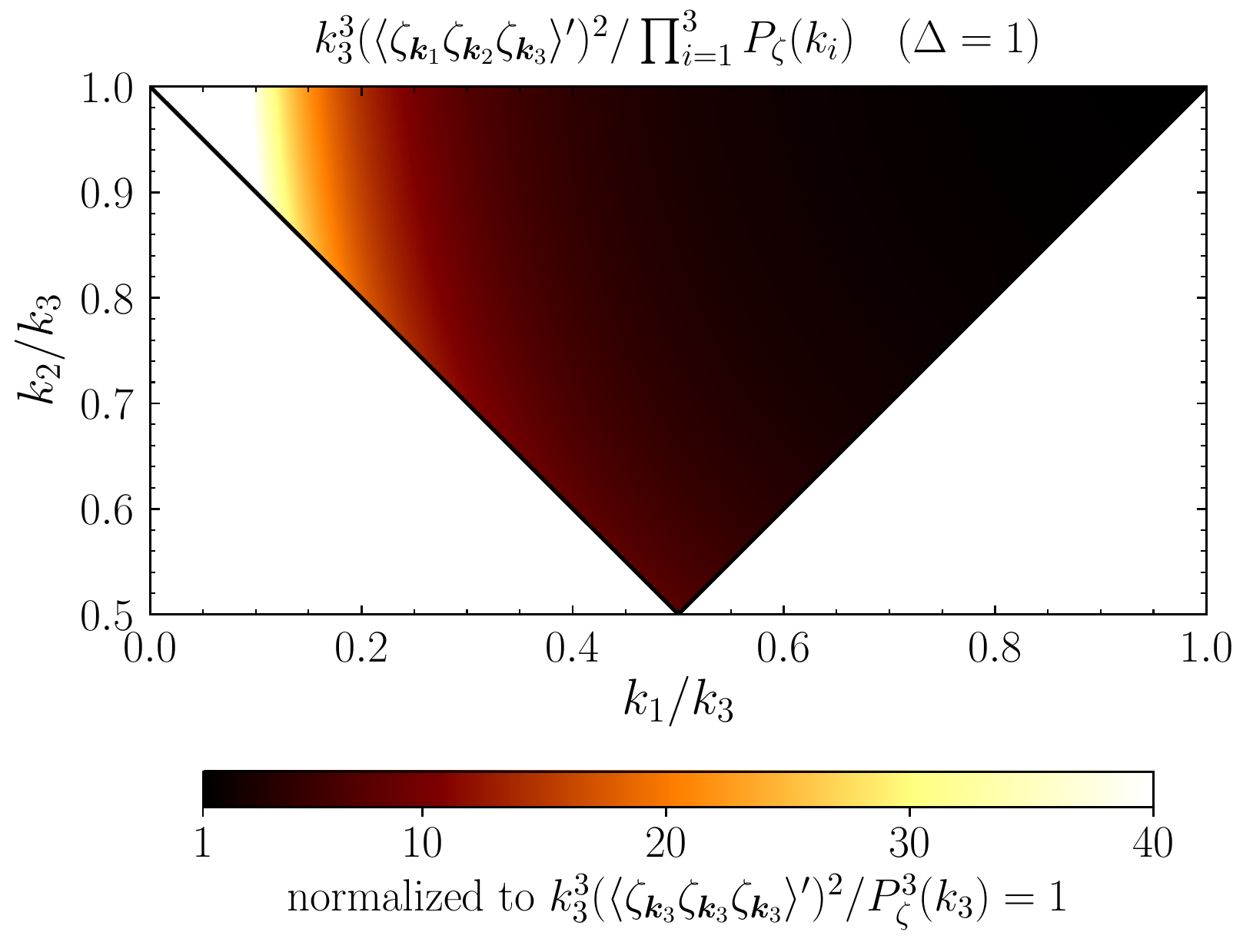}
\end{tabular}
\caption{Plot of $k^3_3({\avg{\zeta_{\k_1} \zeta_{\k_2} \zeta_{\k_3}}'})^2/\prod_{i=1}^3P_\zeta(k_i)$ for $\Delta = 1/2,1$. 
The scaling in the squeezed limit is given by Eqs.~\eqref{eq:bispectrum_templ},~\eqref{eq:def_b_k1_k2_k3}, and is $\sim x_1^{2\Delta - 3}$.} 
\label{shapes_plot}
\end{figure}

\noindent In this appendix we confirm that, as expected, the signal-to-noise for the bispectrum template of Eq.~\eqref{bispectrum_templ_full}, \ie~
\be
\avg{\zeta_{\k_1} \zeta_{\k_2} \zeta_{\k_3}} = (2\pi)^3 \delta(\k_{123})\ \F_s \(\frac{k_1 k_2 k_3}{k_{\rm t}^3/8}\)^{\Delta} P_\zeta(k_1) 
P_\zeta(k_2) \LP_s(\hat{\vec{k}}_1 \cdot \hat{\vec{k}}_2) + \text{$2$ perms.} \,\,,
\ee
peaks in the squeezed limit for all values of $\Delta$ between $0$ and $1$. The relevant quantity for the computation of the Fisher matrix 
is the ratio between the square of the bispectrum $(\avg{\zeta_{\k_1} \zeta_{\k_2} \zeta_{\k_3}}')^2$ and 
$P_\zeta(k_1)P_\zeta(k_2)P_\zeta(k_3)$ (see \eg~Eq.~\eqref{StN_bis}). We then plot this quantity 
as a function of $x_1 = k_1/k_3$ and $x_2 = k_2/k_3$ for $\Delta = 1/2$ and $\Delta=1$, organizing the momenta such that $k_1\leq k_2\leq k_3$ 
and fixing $A_{\rm s}=1$, $n_{\rm s}=1$ and ${\cal F}_s$ such that $k^3_3({\avg{\zeta_{\k_3} \zeta_{\k_3} \zeta_{\k_3}}'})^2P_\zeta^3(k_3) = 1$. 
This is shown in Fig.~\ref{shapes_plot} for $s=0$. We see that, indeed, the signal-to-noise peaks for $x_1\to 0$ and $x_2\to 1$. 
For a general value of $\Delta$, in this limit we see that this ratio goes as $x_1^{2\Delta-3}$. 
For $0\leq\Delta\leq 1$ we are then sure that most of the signal comes from the squeezed limit.

\section{A separable template for the trispectrum}
\label{appendix-B}

In this appendix we show that one can neglect the $x$- and $y$-dependence of $F_L(\Delta, x,y)$ in the formula for the reduced trispectrum, Eq.~\eqref{reduced_trispectrum}. 
The function $F_L (\Delta, x,y)$ can be explicitly evaluated for a given value of $\Delta$. For instance, if $\Delta=0$ it reads
\be
\begin{split}
F_L(0,x,y) = 2^{n_{\rm s}-2} \pi ^2 \, A_{\rm s} \, k_*^{n_{\rm s}-1} \, x^{n_{\rm s}+1} \(\frac{y}{x}\)^L &\ \frac{\Gamma \(L+{n_{\rm s}}/{2}-{1}/{2}\)}{\Gamma \(2-{n_{\rm s}}/{2}\)} \,\times \\ 
&{\ _2\widetilde{F}_1}\({n_{\rm s}}/{2}-1,L+n_{\rm s}/2-1/2; \, L+{3}/{2}; \,({y}/{x)^2}\)\,\,.
\end{split}
\ee
From the above expression it is easy to see that 
\be
\Delta r^2 \, \left.\frac{\d_x\d_y F(0,x,y)}{F(0,x,y)}\right|_{x,y=r_*} = -\frac{L+L^2}{n_{\rm s}-1}\ \frac{\Delta r^2}{r_*^2} + \mathcal{O} (n_{\rm s}-1)\,\,,
\ee
where $\Delta r_*$ is the width of the peak around the recombination distance of the function $\alpha_\ell(x)$. 
Typically $(\Delta r_*/r_*)^2 \lesssim\num{2d-4}$, therefore the approximation $F_L(0,x,y) \simeq F_L(0,r_*,r_*)$ is well justified. 
Following the same steps we did for $\Delta=0$ one can show that this approximation holds also for $\Delta = 1/2,1$ and, 
more in general, for every value of $\Delta$ in the range $0\leq\Delta\leq1$. 

\section{Sachs-Wolfe approximation}
\label{appendix-C}

{The Sachs-Wolfe (SW) approximation is useful to compute the signal-to-noise of Eqs.~\eqref{StN_bis},~\eqref{StN_tris} without needing to perform the integrals 
in the definitions of Eqs.~\eqref{f_l_xy},~\eqref{alpha_ell},~\eqref{beta_ell} numerically.} 

Neglecting acoustic physics (together with Doppler and ISW effects) 
consists in assuming $\Delta_\ell(k) = - j_\ell(k r_*)/5$, with $j_\ell$ a spherical Bessel function. 
This is a good approximation of the transfer functions for $\ell \lesssim 100$: however, as one can see in the plots of Fig.~\ref{fig_signal_to_noises}, 
the corresponding result well approximates the exact Fisher matrix even at higher multipoles (see also \cite{Babich:2004yc}). 

With this approximation the expression for $\alpha_\ell$, \ie~Eq.~\eqref{alpha_ell}, is greatly simplified. $\alpha_\ell$ takes the form
\be
\alpha_\ell(x) = {-\frac{\delta(r_\ast-x)}{5 \sqrt{r_*^3 \, x}}}\,\,.
\ee
Consequently, the integrals of Eqs.~\eqref{bis_integral},~\eqref{tris_integral} become
\begin{align}
{R_{\rm bis}}\,^{\ell_1 \ell_2 \ell_3}_{L_1 L_2} (\Delta) &= {-\frac{1}{5}} \, \beta_{\ell_1 L_1}(\Delta,r_*) \beta_{\ell_2 L_2}(-\Delta,r_*)\,\,, \\
R_{\rm tris}^{\ell_1 L_1\ell_2}(\Delta) &= {- \frac{1}{5}} \, \beta_{\ell_1L_1}(-\Delta,r_*) \,\,.
\end{align}
To evaluate the functions $\beta_{\ell L}$ and $F_L$ we further assume a scale-invariant primordial power spectrum, $P_\zeta (k) = 2\pi^2 A_{\rm s}/k^3$. 
Therefore, for the massless case we get
\begin{align}
\beta_{\ell L} (0,r_*) &= {- \frac{\pi^2}{10}} \, A_{\rm s} \, \frac{\Gamma\(\frac{\ell+L}{2}\)}{\Gamma\(\frac{\ell-L+3}{2}\) 
\Gamma\(\frac{L-\ell+3}{2}\) \Gamma\(\frac{\ell + L }{2}+2\)}\,\,, \label{alpha_SW} \\
F_L(0,r_*, r_*) &=25\, C^{\rm SW}_L\,\,,
\end{align}
where $C_\ell^{\rm SW}=(2\pi/25)\times A_{\rm s}/\ell(\ell+1)$ is the angular power spectrum in the SW approximation. 
For $\Delta=1/2$, the expressions of $\beta_{\ell L}$ and $F_L$ are
\begin{align}
\beta_{\ell L}(1/2,r_*) &= - \frac{\pi^{5/2} A_{\rm s}}{10\sqrt{2 r_*}} \, 
\frac{\Gamma\( \frac{2 \ell + 2L + 1}{4}\)}{\Gamma\(\frac{2\ell - 2L +5}{4}\) \Gamma\(\frac{2L-2\ell+5}{4}\) \Gamma\( \frac{2\ell + 2L + 7}{4}\)} \,\,,\\
\beta_{\ell L}(-1/2,r_*) &= - \frac{3 \pi^{5/2} A_{\rm s} \sqrt{r_*}}{40\sqrt{2}} \, 
\frac{\Gamma\( \frac{2 \ell + 2L -1}{4}\)}{\Gamma\(\frac{2\ell - 2L +7}{4}\) \Gamma\(\frac{2L-2\ell+7}{4}\) \Gamma\( \frac{2\ell + 2L + 9}{4}\)}\,\,, \\
F_L(1/2, r_*, r_*) &= \frac{2 \pi^2 A_{\rm s} }{r_*(2L +1)}\,\,.
\end{align}
Finally, for $\Delta = 1$, the expressions of $\beta_{\ell L}$ and $F_L$ are
\begin{align}
\beta_{\ell L}(1,r_*) &= -\frac{\pi^2 A_{\rm s}}{5 r_*} \,\frac{\Gamma\( \frac{\ell+L+1}{2} \)}{\Gamma \( \frac{\ell-L+2}{2} \) \Gamma\(\frac{-\ell +L +2}{2} \) \Gamma\(\frac{\ell+L+3}{2}\)}\,\,, \\
\beta_{\ell L}(-1,r_*) &= \frac{\pi^2 A_{\rm s} \, r_*}{10} \, \frac{\Gamma\( \frac{\ell+L-1}{2} \)}{\Gamma \( \frac{\ell-L+4}{2} \) 
\Gamma\(\frac{-\ell +L +4}{2} \) \Gamma\(\frac{\ell+L+5}{2}\)} \,\,, \\
F_L(1, r_*, r_*) &= \pi^{3/2} A_{\rm s} \, \frac{k_*^{1-n_{\rm s}}}{r_*^{1+n_{\rm s}}} \, 
\frac{\Gamma\(\frac{1-n_{\rm s}}{2}\) \Gamma\( \frac{2L+1+n_{\rm s}}{2}\)}{\Gamma\(\frac{2-n_{\rm s}}{2}\) \Gamma\( \frac{2L+3-n_{\rm s}}{2}\)}\,\,.
\end{align}
Notice that in the last line we reintroduced the scale dependence of the primordial power spectrum: $P_\zeta (k) = 2\pi^2 A_{\rm s}/k^3 \times (k/k_\ast)^{n_{\rm s}-1}$. 
This is because, in the scale invariant limit, Eq.~\eqref{f_l_xy} formally diverges for $\Delta=1$.

\section{Cosine with the standard bispectrum templates}
\label{appendix-D}

\begin{table}[h]
\centering
\caption[.]{Cosine between the bispectrum template of Eq.~\eqref{bis_temp} for $\Delta=1$ 
and the local, equilateral and orthogonal templates.} 
\label{cosine_table}
\centering
\medskip
\begin{tabular}{lccc}
\toprule
${\cal C}$ & $s=0$ & $s=2$ & $s=4$ \\
\midrule
local & 0.68 & 0.26 & 0.28 \\[2ex]
equilateral & 0.82 & 0.92 & 0.88 \\[2ex]
orthogonal & -0.39 & 0.41 & 0.38 \\
\bottomrule
\end{tabular}
\end{table}

\noindent In this appendix we collect the values of the cosine between the bispectrum template of Eq.~\eqref{bis_temp} and the local, equilateral and orthogonal templates. 
Following \cite{Babich:2004gb}, the cosine is computed as (assuming a scale-invariant power spectrum)
\begin{equation}
\label{eq:cosine-A}
{\cal C}({\cal S}_i,{\cal S}_j)=\frac{{\cal S}_i\cdot{\cal S}_j}{\sqrt{{\cal S}_i\cdot{\cal S}_i}\,\sqrt{{\cal S}_j\cdot{\cal S}_j}}\,\,,
\end{equation}
where 
\begin{equation}
\label{eq:cosine-B}
{\cal S}_i\cdot{\cal S}_j=\int_{\cal V}{\rm d} x_1{\rm d} x_2\,{\cal S}(x_1,x_2,1)\,{\cal S}_j(x_1,x_2,1)\,\,.
\end{equation}
The shape function $\cal S$ is defined by
\begin{equation}
\label{eq:cosine-C}
{\cal S}(k_1,k_2,k_3)=(k_1k_2k_3)^2\avg{\zeta_{\k_1} \zeta_{\k_2} \zeta_{\k_3}}'\,\,,
\end{equation}
while $\cal V$ is the set $\num{d-3}\leq x_1\leq 1-\num{d-3}, 1-x_1\leq x_2\leq 1-\num{d-3}$. The value \num{d-3} has been chosen because 
it is roughly the ratio between the longest and shortest scales that we can access in the CMB. 

The values of the cosine for $\Delta=1$ are reported in Tab.~\ref{cosine_table}. We see that 
there is a sizable overlap only with the equilateral template, while the cosine with the local and orthogonal templates is always small for all 
the values of the spin we considered in this work.

\bibliographystyle{utphys}\bibliography{my_biblio}

\end{document}